\documentclass[twocolumn]{aastex63}

\usepackage{multirow}
\usepackage{graphicx}

\received{2022 May 20}
\accepted{2023 February 24}
\submitjournal{ApJ}

\shorttitle{UV \& Ly$\alpha$ halos of LAEs across environments at $z=2.84$}
\shortauthors{Kikuta et al.}

\begin{document}
\title{UV \& Ly$\alpha$ halos of Ly$\alpha$ emitters across environments at $z=2.84$\footnote{This research is based on data collected at the Subaru Telescope, which is operated by the National Astronomical Observatory of Japan.}}

\correspondingauthor{Satoshi Kikuta}
\email{satoshi.kikuta@nao.ac.jp}

\author[0000-0003-3214-9128]{Satoshi Kikuta}
\affiliation{National Astronomical Observatory of Japan, 2-21-1, Osawa, Mitaka, Tokyo, 181-8588, Japan}
\affiliation{Center for Computational Sciences, University of Tsukuba,
1-1-1, Ten-nodai, Tsukuba, Ibaraki, 305-8577, Japan}

\author[0000-0003-1747-2891]{Yuichi Matsuda}
\affiliation{Graduate University for Advanced Sciences (SOKENDAI),
2-21-1, Osawa, Mitaka, Tokyo, 181-8588, Japan}
\affiliation{National Astronomical Observatory of Japan,
2-21-1, Osawa, Mitaka, Tokyo, 181-8588, Japan}

\author[0000-0001-8819-6877]{Shigeki Inoue}
\affiliation{Department of Cosmosciences, Graduates School of Science, Hokkaido University, Sapporo, Hokkaido 060-0810, Japan}

\author[0000-0002-4834-7260]{Charles C. Steidel}
\affiliation{California Institute of Technology, MS 249-17, Pasadena, CA 91125, USA}

\author[0000-0001-8531-9536]{Renyue Cen}
\affiliation{School of Physics, Zhejiang University, Hangzhou, 310058, China}

\author[0000-0003-1887-6732]{Zheng Zheng}
\affiliation{Department of Physics and Astronomy, University of Utah, 115 South 1400 East, Salt Lake City, UT 84112, USA}

\author{Hidenobu Yajima}
\affiliation{Center for Computational Sciences, University of Tsukuba,
1-1-1, Ten-nodai, Tsukuba, Ibaraki, 305-8577, Japan}

\author[0000-0002-8857-2905]{Rieko Momose}
\affiliation{Carnegie Observatories, 813 Santa Barbara Street, Pasadena, CA 91101, USA}
\affiliation{Department of Astronomy, School of Science, The University of Tokyo, 7-3-1 Hongo, Bunkyo-ku, Tokyo, 113-0033, Japan}

\author[0000-0001-6186-8792]{Masatoshi Imanishi}
\affiliation{Graduate University for Advanced Sciences (SOKENDAI),
2-21-1, Osawa, Mitaka, Tokyo, 181-8588, Japan}
\affiliation{National Astronomical Observatory of Japan, 
2-21-1, Osawa, Mitaka, Tokyo, 181-8588, Japan}

\author[0000-0002-3852-6329]{Yutaka Komiyama}
\affiliation{Department of Advanced Sciences, Faculty of Science and Engineering, Hosei University,
3-7-2 Kajino-cho, Koganei-shi, Tokyo 184-8584, Japan}

\begin{abstract}
We present UV \& Ly$\alpha$ radial surface brightness (SB) profiles of Ly$\alpha$ emitters (LAEs) at $z=2.84$ detected with the Hyper Suprime-Cam (HSC) on the Subaru Telescope. 
The depth of our data, together with the wide field coverage including a protocluster, enable us to study the dependence of Ly$\alpha$ halos (LAHs) on various galaxy properties, including Mpc-scale environments. 
UV and Ly$\alpha$ images of 3490 LAEs are extracted, and stacking the images yields SB sensitivity of $\sim1\times10^{-20}\mathrm{~erg~s^{-1}~cm^{-2}~arcsec^{-2}}$ in Ly$\alpha$, reaching the expected level of optically thick gas illuminated by the UV background at $z\sim3$. Fitting of the two-component exponential function gives the scale-lengths of $1.56\pm0.01$ and $10.4\pm0.3$ pkpc. 
Dividing the sample according to their photometric properties, we find that while the dependence of halo scale-length on environment outside of the protocluster core is not clear, LAEs in the central regions of protoclusters appear to have very large LAHs which could be caused by combined effects of source overlapping and diffuse Ly$\alpha$ emission from cool intergalactic gas permeating the forming protocluster core irradiated by active members.
For the first time, we identify ``UV halos'' around bright LAEs which are probably due to a few lower-mass satellite galaxies. 
Through comparison with recent numerical simulations, we conclude that, while scattered Ly$\alpha$ photons from the host galaxies are dominant, star formation in satellites evidently contributes to LAHs, and that fluorescent Ly$\alpha$ emission may be boosted within protocluster cores at cosmic noon and/or near bright QSOs.
\end{abstract}


\section{Introduction} \label{sec:intro}

Gas inflow from the cosmic web fuels star formation and active galactic nucleus (AGN) activity in galaxies. Following these activities, outflows play a significant role in expelling or heating a large amount of gas, thereby reducing further star formation and supermassive black hole (SMBH) growth. 
The circumgalactic medium (CGM) contains vital information on these flow components and thus holds an important key to revealing galaxy evolution \citep[see][for a recent review]{Tumlinson2017}. 
The CGM of $z\gtrsim2$ star-forming galaxies is now routinely detected as diffuse Ly$\alpha$ nebulae, or Ly$\alpha$ halos (LAHs), around star-forming galaxies such as Ly$\alpha$ emitters (LAEs) and Lyman break galaxies (LBGs) at high-redshift both individually \citep{Rauch2008,Wisotzki2016, Leclercq2017,Erb2018,Bacon2021,Kusakabe2022} and through a stacking technique \citep{Hayashino2004,Steidel2011,Matsuda2012,Feldmeier2013,Momose2014,Momose2016,Xue2017,LujanNiemeyer2022,LujanNiemeyer2022a}.
Together with a technique based on absorption lines in the spectra of neighboring background sources \citep{Adelberger2005,Steidel2010,Rudie2012,Chen2020igm,Muzahid2021}, LAHs have provided a crucial empirical window into the CGM of distant galaxies.

To extract useful information on the CGM from LAHs, the physical origins of the 
Ly$\alpha$ emission should be identified. 
Ly$\alpha$ surface brightness (SB) profiles of LAHs hold the key since they are determined by the distribution and kinematics of gas and the relative importance of various Ly$\alpha$ production mechanisms such as scattering of Ly$\alpha$ photons from host galaxies, star formation in neighboring galaxies, collisional excitation of inflow gas powered by gravitational energy (sometimes called gravitational cooling radiation), and recombination following photoionization by external sources, often referred to as ``fluorescence.'' 
Theoretical studies have attempted to reproduce and predict observed LAHs by considering these mechanisms \citep{haiman2001,Dijkstra2009blob,Goerdt2010,Kollmeier2010,Faucher-Giguere2010,Zheng2011,Dijkstra2012halo,Rosdahl2012,Yajima2013,Cen2013,Cantalupo2014,Lake2015,Mas-Ribas2016,Gronke2017, Mitchell2020a,Byrohl2021}. 
Powerful outflows \citep[so-called ``superwind'',][]{Taniguchi2000,Mori2004}
have also been proposed to excite gas, but often for more energetic/massive counterparts such as Ly$\alpha$ blobs (LABs) and nebulae around QSOs and radio galaxies\footnote{There is no clear demarcation, but conventionally LABs refer to extended Ly$\alpha$ nebulae that are particularly bright ($L_\mathrm{Ly\alpha}>10^{43}\mathrm{~erg~s^{-1}}$) but without obvious AGN activity at optical wavelengths.}.

Dependence of LAH shapes on e.g., their hosts' halo mass and large-scale overdensity is naturally expected because both gas and sources of Ly$\alpha$ and ionizing photons are more abundant in massive halos and/or denser environments \citep{Zheng2011,Mas-Ribas2016, Kakiichi2018}. 
Current simulations cannot treat all relevant physics with sufficient accuracy and it is only very recently that such predictions are reported in the literature with a statistical number of simulated galaxies \citep[e.g.,][]{Byrohl2021}. 
Observations of LAHs can help theorists pin down which Ly$\alpha$ production processes are at work by revealing the dependence of LAH SB profile shapes on various properties such as UV and Ly$\alpha$ luminosity, Ly$\alpha$ equivalent width (EW$_\mathrm{Ly\alpha,0}$, \citealt{Momose2014,Momose2016,Wisotzki2016,Wisotzki2018,Leclercq2017}), and the large-scale environment \citep{Matsuda2012,Xue2017}. 
The results in the literature are, however, far from converging (see e.g., Figure 12 of \citet{Leclercq2017}). 
It is often parametrized by an exponential function $\propto \exp(-r/r_\mathrm{h})$ with a scale-length $r_\mathrm{h}$, which is fit to observed profiles.
The reported scale-lengths of individual LAEs as a function of UV magnitude show a large scatter (from $<1$ physical kpc (pkpc hereafter) to $\sim10$ pkpc) and the relation for stacked LAEs shows large differences as well.
In the case of large-scale environment, relevant observations of LAHs of LAEs are still scarce. 
First, \citet{Steidel2011} found very large LAHs with a scale-length of 25 pkpc around LBGs in three protoclusters at $z=2.3$-3.1. 
Following this result, \citet{Matsuda2012} suggested that the scale-length of LAHs of LAEs are proportional to galaxy overdensity squared $\delta^2$. 
On the other hand, \citet{Xue2017} found no such dependence with LAEs in two overdense regions at $z=2.66$ and $z=3.78$.

A major problem with some previous work is poor sensitivity. While only a few studies investigated LAHs with deep images of a fairly large sample of LAEs ($N_\mathrm{LAE}>2000$; \citealt{Matsuda2012,Momose2014}), others used the insufficient number of LAEs ($N_\mathrm{LAE}\sim$ a few$\times100$--1000) and/or images taken with 4m telescopes \citep[e.g.,][]{Feldmeier2013,Xue2017}. Because LAHs beyond the virial radii of LAEs have extremely low SB ($<10^{-19} \mathrm{~erg~s^{-1}~cm^{-2}~arcsec^{-2}}$), interpretation of relations derived from shallow data would not be straightforward. 
Even with sufficiently deep data, the extent of LAHs is difficult to measure. 
Disagreements among different studies can be attributed in part to different fitting methods, fitting range, radial bin size of SB profiles, and sensitivity of observational data among each study. 
To alter this situation and to provide a firm observational basis for theorists, a well-controlled statistical sample of LAEs drawn from a wide dynamic range of environments, with sufficiently deep images, is required. 

In this paper, we present a new LAH study with deep narrow-band (NB468) data obtained using the Hyper Suprime-Cam \citep[HSC;][]{Miyazaki2012a} on the Subaru Telescope toward the HS1549 protocluster at $z=2.84$ \citep{Trainor2012,Mostardi2013,Kikuta2019} to probe what shapes LAHs. 
Thanks to the HSC's large field of view ($\phi\sim1.5$ deg, corresponding to 160 comoving Mpc at $z=2.84$), we can construct a large LAE sample across environments from a protocluster to surrounding lower density fields at the same time. 
The sample size of our study of $N=3490$ is one of the largest to date, giving robust UV and Ly$\alpha$ SB profiles to be compared with simulations. 
As a result, we for the first time detect ``UV halos'' which directly prove the contribution of star formation in satellite galaxies. 
Moreover, we detect very extended LAHs for the protocluster LAEs which suggest an important role of locally enhanced ionizing radiation fields for LAHs. 
This paper is structured as follows. In Section \ref{sec:data}, we describe our LAE sample, followed by how we divide them for the stacking analyses described in Section \ref{sec:analyses}. The results of the analyses are shown in Section \ref{sec:result}. Based on these, we present discussion in Section \ref{sec:discussion} and summarize the work in Section \ref{sec:summary}.
Throughout this paper, we use the AB magnitude system and assume a cosmology with $\Omega_\mathrm{m} = 0.3$, $\Omega_\Lambda= 0.7$, and $H_0 = 70 \mathrm{~km~s^{-1}~Mpc^{-1}}$, unless otherwise noted.

\section{Data and sample} \label{sec:data}

\begin{figure*}
\plotone{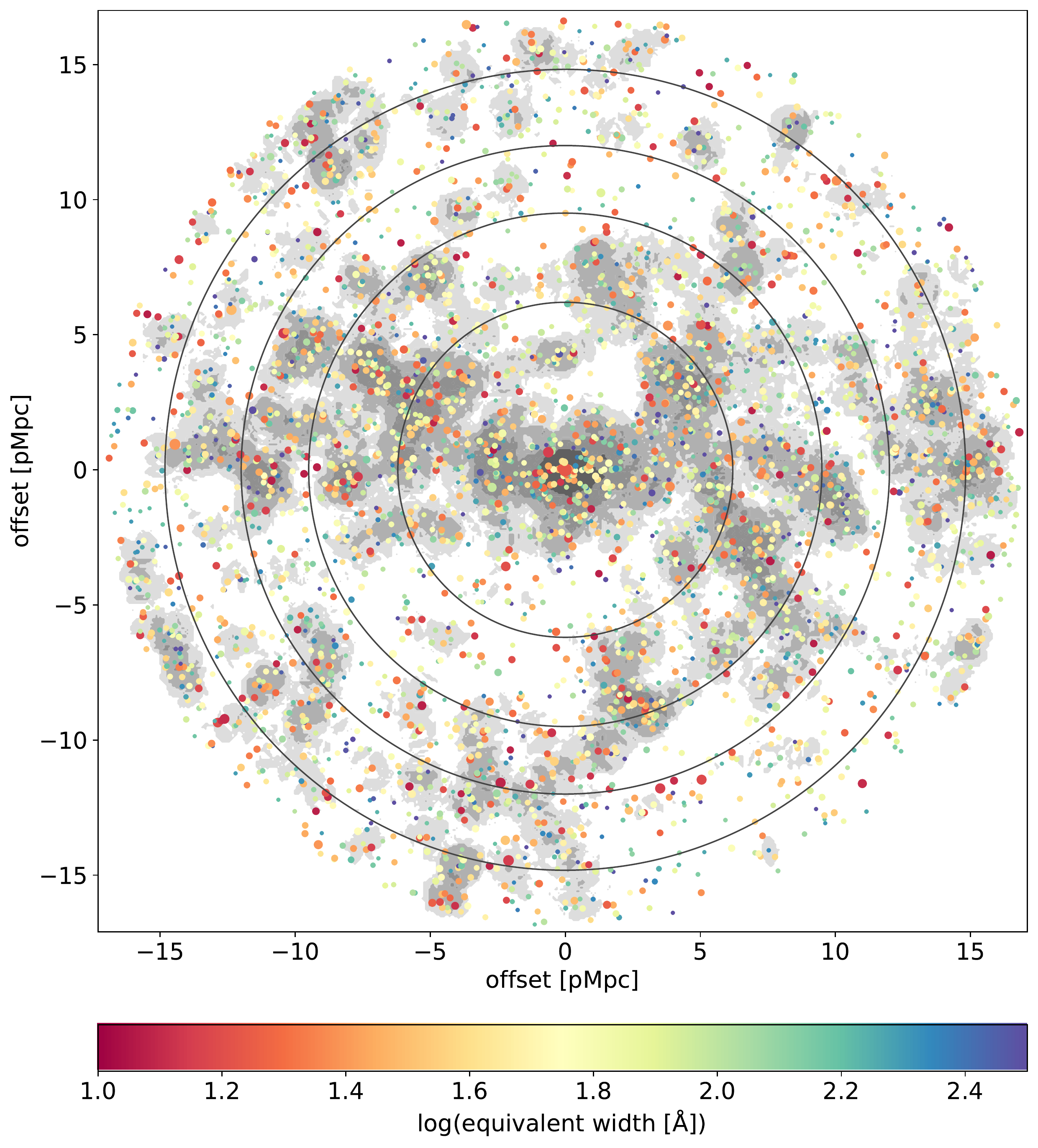}
\caption{
  Sky distribution of our LAEs (colored dots). The HLQSO is located at the origin (0,0). North is up and east to the left. The colors indicate their rest-frame Ly$\alpha$ equivalent width and the sizes indicate their UV absolute magnitude (the larger, the brighter). Gray contours indicate their overdensity $\delta$, with each level showing a different group defined in Table \ref{tab:halostack}. 
  Concentric black circles indicate radii with which the distance subsamples are defined.
}
\label{fig:laemap}
\end{figure*}

We used the LAE sample described by \citet{Kikuta2019}. 
Here, we highlight key properties of the data and refer readers to the aforementioned paper for details. 
The target protocluster contains a hyperluminous QSO, namely HS1549+1919 ($L_\mathrm{1450\AA}=1.5\times10^{14}~L_\odot$, \citealt{Trainor2012,Mostardi2013}) at its center. 
The field was observed in the g-band (central wavelength $\lambda_\mathrm{c}=4712$\AA, $\mathrm{FWHM}=1479$\AA) and NB468 ($\lambda_\mathrm{c}=4683$\AA, $\mathrm{FWHM}=88$\AA) narrow-band filters. 
The global sky subtraction method\footnote{https://hsc.mtk.nao.ac.jp/pipedoc\_e/e\_tips/skysub.html\#global-sky} 
was used to estimate and subtract the sky on scales larger than that of individual CCDs in the mosaic, with a grid size of 6000 pixels (17\arcmin) not to subtract diffuse emission. 
The FWHMs of stellar sources in the final images are 0{\mbox{$.\!\!\arcsec$}}77 (0{\mbox{$.\!\!\arcsec$}}65) for g-band (NB468). 
The NB468 image was smoothed with a circular Gaussian function to match the FWHM of stellar sources in the g-band image. 
A 5$\sigma$ limiting magnitude measured with 1{\mbox{$.\!\!\arcsec$}}5 diameter aperture is 27.4 (26.6) mag for the g-band (NB468) image. 
Our criterion for NB excess, $\mathrm{g} - \mathrm{NB468}>0.5$, corresponds to $\mathrm{EW_{obs}}>45$ \AA~ after considering the 0.1 mag offset. 
SExtractor \citep{Bertin1996} was used to perform 1{\mbox{$.\!\!\arcsec$}}5 aperture photometry with double-image mode, using the NB468 image as the detection band and a background mesh size of 64 pixels ($=11$\arcsec) for local sky estimation.
This small mesh size is only used for LAE detection, since this is optimal for detecting compact sources such as distant galaxies.
As a result, we detected 3490 LAEs within $r<36\arcmin$ from the QSO position. Their sky distribution is shown in Figure \ref{fig:laemap}.

To study the dependence of LAHs on various photometric galaxy properties, we divided the sample into several groups according to the following five quantities: UV magnitude, Ly$\alpha$ luminosity, rest-frame Ly$\alpha$ equivalent width (EW$_0$), environment, and distance from the HLQSO, as summarized in Table \ref{tab:halostack}. 
The first three quantities, which are derived with 1{\mbox{$.\!\!\arcsec$}}5 aperture ($=9$ pixels $=2\times\mathrm{[PSF~FWHM]}$), are obviously not independent of one another.
UV continuum and Ly$\alpha$ luminosity are both good proxies for star formation rate (SFR), but the latter's resonant nature and a different impact of dust extinction thereby make interpretation difficult \citep[e.g.,][]{Scarlata2009}. 
Moreover, UV slope or hardness of UV emission is a strong function of age and metallicity, and thus the Ly$\alpha$ equivalent width changes accordingly \citep{Schaerer2003a,Hashimoto2016}. 
There is a known observational relation between the EW distribution and UV limiting magnitude, known as ``the Ando effect'' \citep[a fainter UV threshold tends to include more high-EW LAEs;][]{Ando2006}\footnote{Note that this effect could be purely attributed to the selection bias and not intrinsic \citep{Nilsson2009a,Zheng2010}.}.
Since Ly$\alpha$ emission can be powered by mechanisms other than star formation in galaxies, UV magnitude is the most robust to use here as a tracer for SFR. 
Binning in UV magnitude, Ly$\alpha$ luminosity, and Ly$\alpha$ equivalent width (as well as distance from the HLQSO) were done so that each subsample has approximately the same number of LAEs ($N_\mathrm{LAE}\sim700$). 

The projected distance from the HLQSO is used to test whether QSO radiation affects the LAHs of surrounding LAEs. HS1549 is so luminous that the entire field covered by the HSC could experience a higher ionizing radiation field than the cosmic average at $z\sim3$ \citep{Haardt2012} if QSO radiation has had time to propagate, and additional ionization induced by the QSO could increase Ly$\alpha$ luminosity of LAEs in the field (see Section \ref{sec:5distance} for discussion). 
Here, we use a projected distance from the QSO. Note, however, that the NB468 filter has an FWHM of $\Delta \lambda =88$\AA\ or $\Delta z=0.075$, corresponding to 19 pMpc width when centered at $z=2.84$.
This brings uncertainty in a line-of-sight distance and therefore also in real (3D) distance.
The boundaries defining the LAE subsamples are indicated by concentric circles in Figure \ref{fig:laemap}.  
Lastly, grouping based on environment is done using projected (surface) LAE overdensity $\delta\equiv(n-\bar n)/\bar n$ measured locally with an aperture radius of 1{\mbox{$.\!\!\arcmin$}}8 ($=0.83$ pMpc) to be consistent with the measurement of \citet{Matsuda2012}. 
Here, $n$ and $\bar n$ are, respectively, the number of LAEs within a circle with a 1{\mbox{$.\!\!\arcmin$}}8 radius centered at the position of interest, and its average over the entire field.
This division is visually illustrated in Figure \ref{fig:laemap} by gray contours (see also Figures 1 and 2 of \citealt{Kikuta2019}). The boundary of $\delta=2.5$ is set to only include protocluster members in the densest subgroup by visual inspection. The next boundary $\delta=1$ is set by a trade-off between tracing sufficiently dense regions and including adequate numbers to allow sufficient S/N in stacks. The remainder are set so as to roughly equalize the number of LAEs in each bin.

\begin{figure*}
\includegraphics[width=\textwidth]{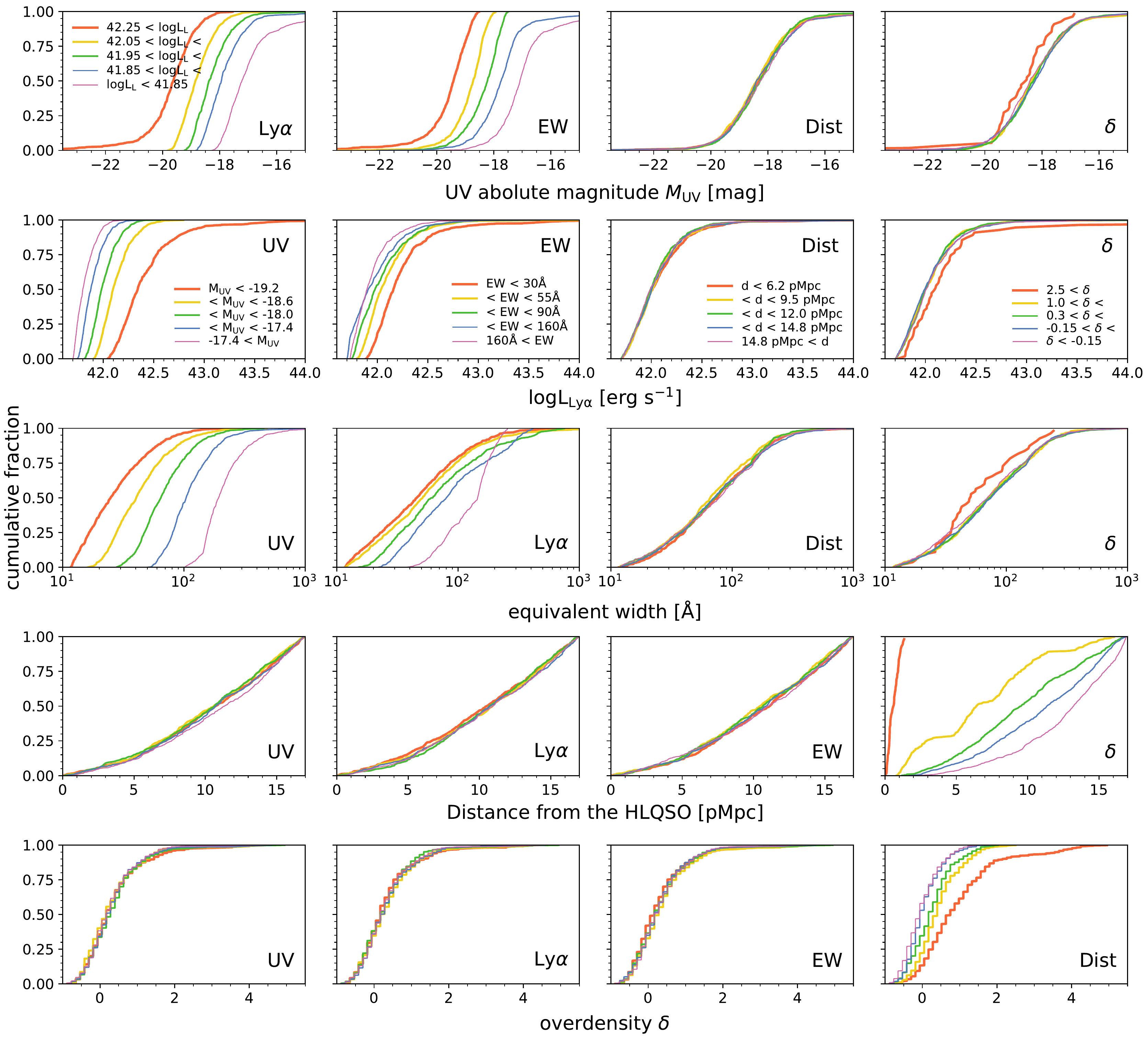}
\caption{
  From top to bottom, cumulative distributions of UV magnitude, Ly$\alpha$ luminosity, rest-frame Ly$\alpha$ EW, distance from the HLQSO, and overdensity of all subsamples described in Table \ref{tab:halostack} and the top-left panel and panels in the second row. In each panel, the quantity used to divide the sample is labeled. Thicker curves present [UV brighter, Ly$\alpha$ brighter, lower EW, farther from the HLQSO, denser] subsamples.
  Note that in the second row, last column, the $2.5<\delta$ subsample contains 4 sources with $L_\mathrm{Ly\alpha}>10^{44}\mathrm{~erg~s^{-1}}$ (including the HLQSO) and thus the red curve does not reach to unity.
}
\label{fig:cumdis0}
\end{figure*}

Figure \ref{fig:cumdis0} shows cumulative distributions of UV and Ly$\alpha$ luminosity, rest-frame Ly$\alpha$ equivalent width, distance from the HLQSO, and overdensity $\delta$ for each division, illustrating correlations between these quantities. 
We note that the odd behavior of the thin blue and purple curves in the second panel from left in the third row of Figure \ref{fig:cumdis0} is likely to be artificial; EWs of LAEs not detected in g band (above $2\sigma$) are just lower limits. 
The samples of faint low-EW LAEs are also incomplete due to the lack of dynamic range in the measurement of $\mathrm{g-NB468}$, distorting the distribution.
The protocluster subsample (LAEs with $\delta>2.5$; thick red curves in panels shown with $\delta$) evidently stands out among others, while the projected distance from the HLQSO seems not to make a significant difference except for $\delta$. These differences should be kept in mind when interpreting the result in Section \ref{sec:result}.
We visualize these quantities also in Figure \ref{fig:laemap} which roughly includes all of above information.

\begin{longrotatetable}
\begin{deluxetable*}{cccccccccc}
\tabletypesize{\scriptsize}
\tablecaption{Subsample used in our stacking analysis and the result of fitting performed in Section \ref{sec:ressub}.}
\tablehead{
\colhead{quantity} & \colhead{criteria} & \colhead{median} & \colhead{$N$} & \colhead{$C_\mathrm{1,Ly\alpha}$} & \colhead{$r_\mathrm{1,Ly\alpha}$} & \colhead{$C_\mathrm{2,Ly\alpha}$} & \colhead{$r_\mathrm{2,Ly\alpha}$} & \colhead{$C_\mathrm{3,Ly\alpha}$} & \colhead{$\alpha$} \\
\colhead{(1)} & \colhead{(2)} & \colhead{(3)} & \colhead{(4)} & \colhead{(5)} & \colhead{(6)} & \colhead{(7)} & \colhead{(8)} & \colhead{(9)} & \colhead{(10)}
}
\startdata
--             &   All    &  --   & 3490 (430) & $2.742 \pm 0.011$  & $1.564 \pm 0.013$  & $1.062 \pm 0.043$  & $10.384 \pm 0.257$  & $1.575 \pm 0.009$  & $2.285 \pm 0.005$  \\ \hline
& $M_\mathrm{UV}<-19.2$ & -19.62 & 690 (0) & $3.782 \pm 0.018$  & $2.106 \pm 0.019$  & $1.369 \pm 0.085$  & $12.795 \pm 0.478$  & $2.058 \pm 0.013$  & $2.057 \pm 0.005$ \\
& $-19.2<M_\mathrm{UV}<-18.6$ & -18.88 & 696 (0) & $2.689 \pm 0.020$  & $1.893 \pm 0.031$  & $0.949 \pm 0.128$  & $10.144 \pm 0.758$  & $1.575 \pm 0.017$  & $2.200 \pm 0.009$ \\
UV magnitude & $-18.6 < M_\mathrm{UV}<-18.0$ & -18.31 & 773 (0) & $2.852 \pm 0.024$  & $1.494 \pm 0.025$  & $0.661 \pm 0.078$  & $11.403 \pm 0.850$  & $1.785 \pm 0.025$  & $2.439 \pm 0.013$ \\
& $-18.0<M_\mathrm{UV}<-17.4$ & -17.73 & 648 (0) & $2.754 \pm 0.031$  & $1.284 \pm 0.028$  & $0.467 \pm 0.059$  & $13.307 \pm 1.166$  & $1.850 \pm 0.037$  & $2.622 \pm 0.020$ \\
& $-17.4<M_\mathrm{UV}$ & -16.92 & 683 (430) & $2.884 \pm 0.034$  & $1.088 \pm 0.027$  & $0.398 \pm 0.048$  & $14.282 \pm 1.259$  & $2.155 \pm 0.053$  & $2.843 \pm 0.026$ \\ \hline
& $42.25<\log L_\mathrm{Ly\alpha}\mathrm{[erg~s^{-1}]}$ & 42.40 & 647 (0) & $6.253 \pm 0.025$  & $1.566 \pm 0.014$  & $2.609 \pm 0.130$  & $9.446 \pm 0.271$  & $3.630 \pm 0.021$  & $2.292 \pm 0.005$ \\
& $42.05<\log L_\mathrm{Ly\alpha}<42.25$ & 42.14 & 833 (5)& $3.382 \pm 0.024$  & $1.484 \pm 0.026$  & $1.423 \pm 0.155$  & $8.092 \pm 0.479$  & $2.092 \pm 0.022$  & $2.392 \pm 0.010$ \\
Ly$\alpha$ luminosity & $41.95<\log L_\mathrm{Ly\alpha}<42.05$ & 41.99 & 610 (26) & $2.726 \pm 0.027$  & $1.495 \pm 0.026$  & $0.558 \pm 0.054$  & $14.929 \pm 1.011$  & $1.610 \pm 0.025$  & $2.383 \pm 0.014$ \\
& $41.85<\log L_\mathrm{Ly\alpha}<41.95$ & 41.90 & 645 (80) & $2.484 \pm 0.030$  & $1.317 \pm 0.034$  & $0.642 \pm 0.096$  & $10.271 \pm 0.966$  & $1.607 \pm 0.032$  & $2.544 \pm 0.020$ \\
& $\log L_\mathrm{Ly\alpha}<41.85$ & 41.79 & 755 (319) & $2.386 \pm 0.032$  & $1.124 \pm 0.030$  & $0.403 \pm 0.044$  & $14.702 \pm 1.182$  & $1.600 \pm 0.039$  & $2.696 \pm 0.025$ \\ \hline
& $12<\mathrm{EW_{0,Ly\alpha}}<30$ \AA & 21.1 \AA & 644 (0) & $2.045 \pm 0.016$  & $2.404 \pm 0.036$  & $0.821 \pm 0.085$  & $13.848 \pm 0.848$  & $1.081 \pm 0.011$  & $1.942 \pm 0.008$ \\
& $30<\mathrm{EW_{0,Ly\alpha}}<55$ \AA & 42.4 \AA & 735 (0) & $2.439 \pm 0.020$  & $1.883 \pm 0.030$  & $1.094 \pm 0.090$  & $11.721 \pm 0.576$  & $1.298 \pm 0.013$  & $2.086 \pm 0.008$ \\
Ly$\alpha$ EW$_0$ & $55<\mathrm{EW_{0,Ly\alpha}}<90$ \AA & 70.5 \AA & 698 (0) & $2.898 \pm 0.026$  & $1.480 \pm 0.031$  & $1.137 \pm 0.137$  & $8.862 \pm 0.609$  & $1.753 \pm 0.023$  & $2.375 \pm 0.012$ \\
& $90<\mathrm{EW_{0,Ly\alpha}}<160$ \AA & 121 \AA & 727 (100) & $3.017 \pm 0.027$  & $1.381 \pm 0.025$  & $0.680 \pm 0.071$  & $11.880 \pm 0.811$  & $1.904 \pm 0.028$  & $2.497 \pm 0.014$ \\
& 160 \AA $<\mathrm{EW_{0,Ly\alpha}}$ & 216 \AA & 686 (330) & $3.433 \pm 0.032$  & $1.178 \pm 0.024$  & $0.646 \pm 0.078$  & $10.951 \pm 0.869$  & $2.450 \pm 0.043$  & $2.734 \pm 0.018$ \\ \hline
& $d_\mathrm{Q} < 6.2$ pMpc & 4.05 pMpc & 679 (81) & $2.709 \pm 0.023$  & $1.659 \pm 0.026$  & $0.991 \pm 0.068$  & $13.178 \pm 0.586$  & $1.418 \pm 0.016$  & $2.161 \pm 0.010$ \\
& $6.2 < d_\mathrm{Q} < 9.5$ pMpc & 7.95 pMpc & 739 (81) & $2.710 \pm 0.024$  & $1.540 \pm 0.034$  & $1.353 \pm 0.158$  & $8.356 \pm 0.533$  & $1.600 \pm 0.020$  & $2.304 \pm 0.011$ \\
Projected distance & $9.5 < d_\mathrm{Q} < 12.0$ pMpc & 10.7 pMpc & 633 (70) & $2.687 \pm 0.029$  & $1.442 \pm 0.043$  & $1.639 \pm 0.248$  & $6.983 \pm 0.537$  & $1.641 \pm 0.023$  & $2.358 \pm 0.013$ \\
from the HLQSO & $12.0 < d_\mathrm{Q} < 14.8$ pMpc & 13.5 pMpc & 778 (111) & $2.838 \pm 0.023$  & $1.563 \pm 0.026$  & $0.876 \pm 0.091$  & $10.690 \pm 0.676$  & $1.688 \pm 0.021$  & $2.341 \pm 0.011$ \\
& $14.8 < d_\mathrm{Q}<16.9$ pMpc & 15.9 pMpc & 661 (87) & $2.766 \pm 0.025$  & $1.567 \pm 0.026$  & $0.856 \pm 0.063$  & $13.593 \pm 0.665$  & $1.490 \pm 0.018$  & $2.236 \pm 0.011$ \\ \hline
& $2.5 < \delta$ & 3.78 & 55 (2) & $2.295 \pm 0.058$  & $2.440 \pm 0.079$  & $0.615 \pm 0.069$  & $43.237 \pm 5.108$  & $0.924 \pm 0.026$  & $1.715 \pm 0.019$ \\
& $1.0 < \delta < 2.5$ & 1.33 & 433 (57) & $2.686 \pm 0.031$  & $1.574 \pm 0.039$  & $1.122 \pm 0.140$  & $10.020 \pm 0.737$  & $1.524 \pm 0.024$  & $2.264 \pm 0.014$ \\
Environment & $0.3 < \delta < 1.0$ & 0.63 & 944 (116) & $2.781 \pm 0.022$  & $1.543 \pm 0.024$  & $1.005 \pm 0.070$  & $11.743 \pm 0.525$  & $1.533 \pm 0.017$  & $2.257 \pm 0.010$ \\
& $-0.15 < \delta < 0.3$ & 0.05 & 1076 (146) & $2.693 \pm 0.021$  & $1.555 \pm 0.026$  & $1.181 \pm 0.102$  & $9.482 \pm 0.483$  & $1.545 \pm 0.016$  & $2.279 \pm 0.009$ \\
& $-1.0 < \delta < -0.15$ & -0.30 & 982 (109) & $2.812 \pm 0.022$  & $1.513 \pm 0.029$  & $1.249 \pm 0.143$  & $8.284 \pm 0.525$  & $1.703 \pm 0.019$  & $2.352 \pm 0.010$ \\
\enddata
\tablecomments{Column (1) and (2): quantity and criteria used to define subsamples, Column (3): median value of each quantity of each subsample, Column (4): the number of LAEs which satisfy the criteria described in the column (2). The numbers in parenthesis are those of g-band undetected sources. Column (5): $C_\mathrm{1,Ly\alpha}$ in units of $10^{-17} \mathrm{~erg~s^{-1}~cm^{-2}~arcsec^{-2}}$, Column (6): $r_\mathrm{1,Ly\alpha}$ in units of physical kpc, Column (7): $C_\mathrm{2,Ly\alpha}$ in units of $10^{-18} \mathrm{~erg~s^{-1}~cm^{-2}~arcsec^{-2}}$, Column (8): $r_\mathrm{2,Ly\alpha}$ in units of physical kpc, Column (9): $C_\mathrm{3,Ly\alpha}$ in units of $10^{-16} \mathrm{~erg~s^{-1}~cm^{-2}~arcsec^{-2}}$, Column (10): power-law index $\alpha$.
Note that UV magnitude, Ly$\alpha$ luminosity, and its EW are derived with 1{\mbox{$.\!\!\arcsec$}}5 aperture. 
The uncertainties of fitting parameters include fitting errors only.
}
\label{tab:halostack}
\end{deluxetable*}
\end{longrotatetable}

\section{Analyses \label{sec:analyses}}
\subsection{Image Stacking}\label{sec:3stack}
Before stacking, additional sky subtraction needs to be made, as the global sky subtraction (see Section \ref{sec:data}) alone is generally insufficient to eliminate the influence of artifacts such as halos of bright stars. 
Since we are interested in diffuse and extended components, we use a background mesh size of 176 pixels ($=30$ arcsec) for additional background evaluation of g-band and NB images using SExtractor and then subtract the sky. 
Then a Ly$\alpha$ (continuum) image was created by subtracting the g-band (NB468) image from the NB468 (g-band) image after scaling by their relative zero points and considering the difference in the filters' transmission curves assuming flat continuum \citep[see Appendix B of][for details]{Mawatari2012}. 
A segmentation image of the continuum image was used for masking. 
The segmentation image is an output of SExtractor and it specifies which pixel is detected as a source. If detected, that pixel has a non-zero integer value that corresponds to the ID number in the output catalog. 
We set DETECT\_MINAREA and DETECT\_THRESH parameters to 5 and 2.0. We confirmed that the overall results do not depend significantly on the choice of the threshold value.

We created cutout Ly$\alpha$ and UV continuum images centered on each LAE. The centers of the LAEs are identified as their centroids in the NB image. The mask is applied to each cutout image. 
If the LAE at the center of the cutout image is detected in the continuum image, masking for the object is turned off so as not to underestimate Ly$\alpha$ and UV continuum emission near the center. Stacking is executed using the IRAF task ``imcombine'' with median/average and with/without sigma clipping to further eliminate unrelated signals. 
Ly$\alpha$ SB profiles of stacked images are measured in a series of annuli with a width of 2 pixels.

\subsection{Uncertainties and Limitations}\label{sec:3anaunc}
\begin{figure}[!h]
\includegraphics[width=\columnwidth]{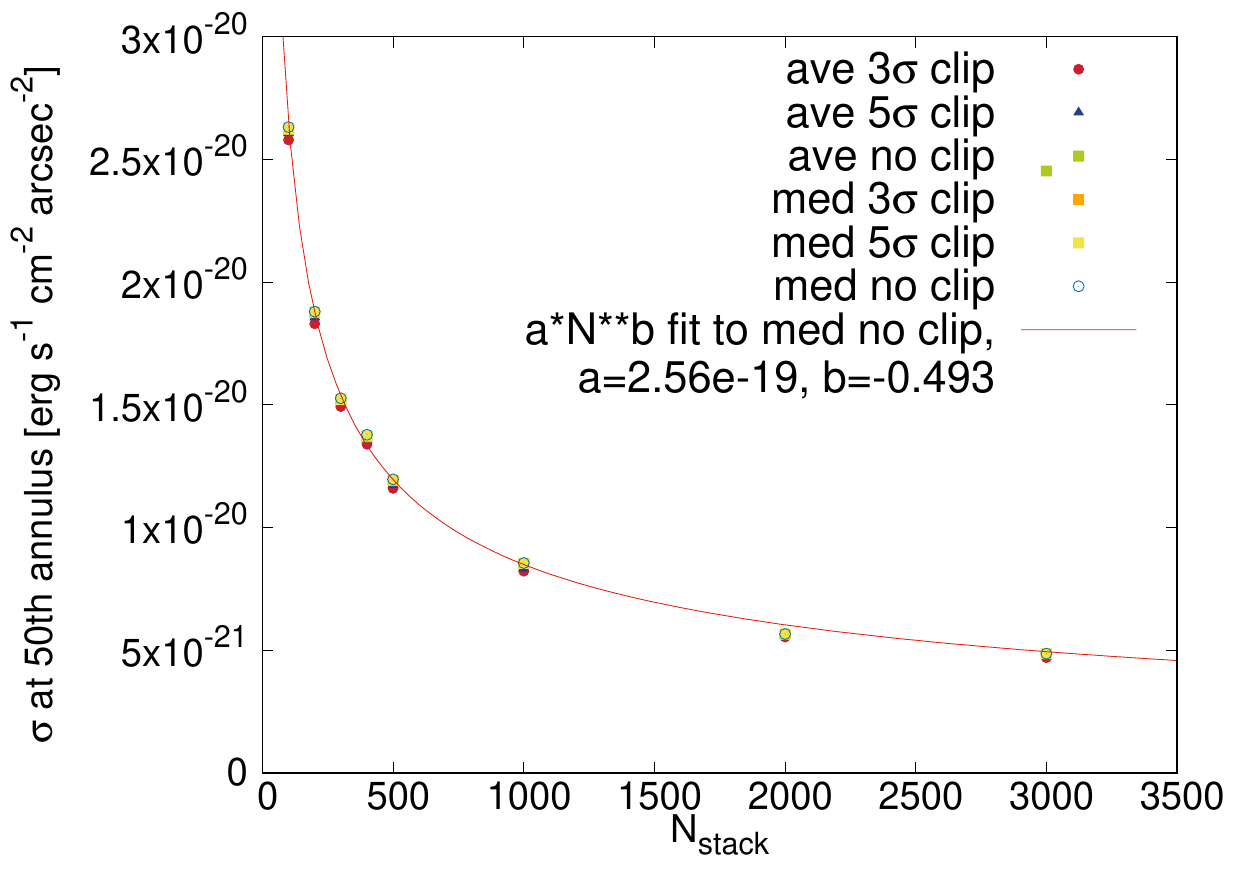}
\includegraphics[width=\columnwidth]{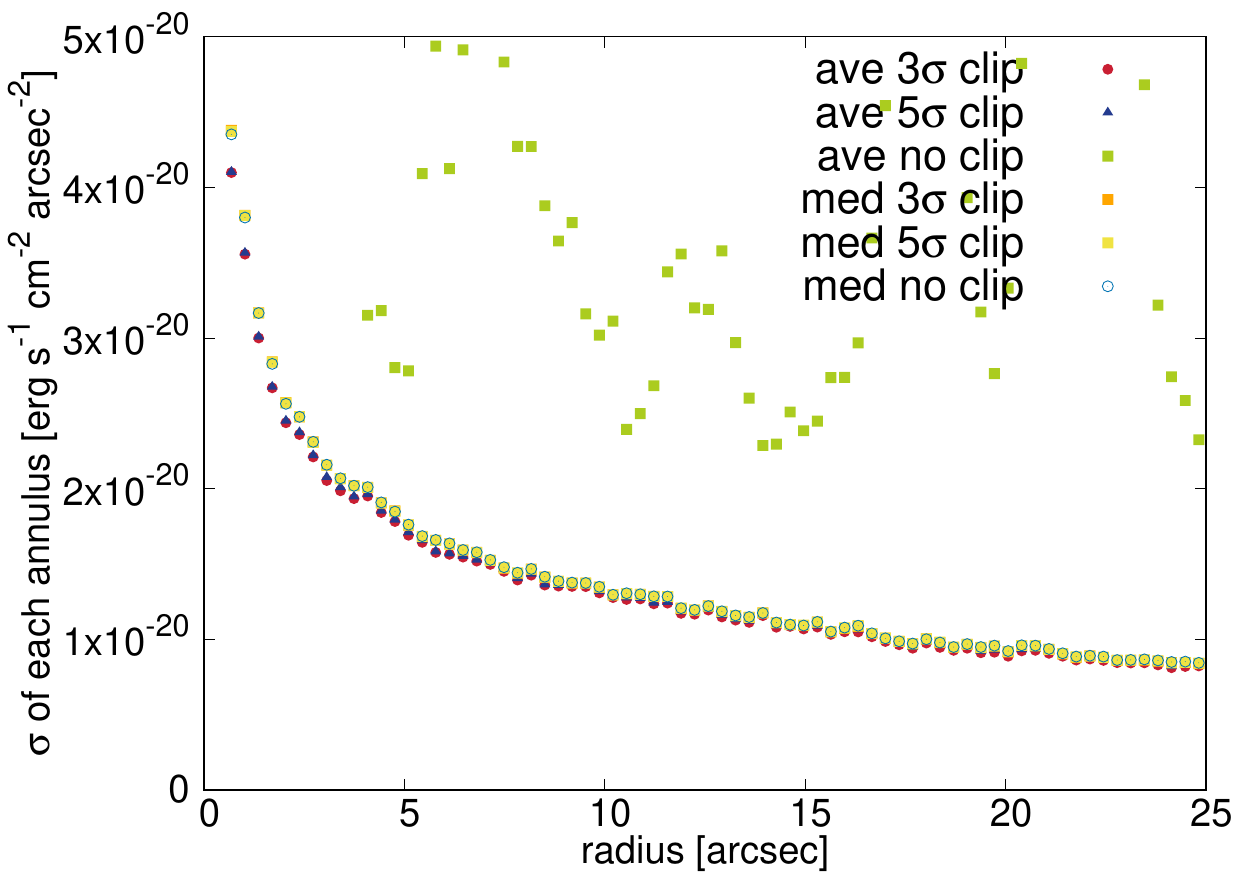}
  \caption{(Top) Estimating the sky noise level of the Ly$\alpha$ image. The x-axis shows how many sky cutouts were stacked, and the y-axis shows the estimated noise level evaluated in 50th annulus from the center (thus $r=$ 100 pixel or 17 arcsec) containing 1270 pixels. Different points indicate the different stacking methods (average VS. median, with 3$\sigma$ or 5$\sigma$ clipping VS. no sigma clipping), which are well converged except for the case with average without sigma clipping. The red curve is a fitting function with a form $a\times N^b$ to the blue circles, which is consistent with inverse square root proportionality ($b=-0.493\pm0.0074$). Symbols for "ave no clip" are mostly out of the upper boundary. 
  (Bottom) The estimated noise level of each annulus for the  $N_\mathrm{stack}=700$ case. Their behavior with respect to the stacking methods is the same as the above figure.
\label{fig:skynoise}}
\end{figure}

To estimate the noise level of the stacked image, we created cutouts of the Ly$\alpha$ and continuum image centered on randomly selected points in the field. After applying the continuum source mask, these ``sky cutouts'' are stacked to make stacked sky images and their radial SB profiles are measured in the same manner as stacked LAE images. This time the turning off of the masking of some sources is not employed. 
The 1$\sigma$ noise level of each annulus is estimated by repeating this 1000 times and deriving the standard deviation of the distribution of total count in the annulus\footnote{When estimating the noise level, using the same aperture shape (in this case, annuli with a width of 2 pixels) is important because spatial correlations between neighboring pixels affect the noise level differently with different aperture shapes.}.
As shown in Figure \ref{fig:skynoise} (Top), we confirmed that the noise level decreases almost as $\propto N^{-1/2}$. 
The result also demonstrates that the choice of stacking method does not make a significant difference in the noise level except for the case of average stacking without sigma clipping (in this case, too many artifacts remain; green points for ``ave no clip'' except for $N=3000$ are far above the graph's upper bound).
Thus, we decided to extrapolate this relation between the noise and $N$ (the number of images for stacking) to estimate the noise level of stacked images with any $N$ rather than to iterate 1000 times for every possible $N$ in the column (4) of Table \ref{tab:halostack}. 
The same method is used to estimate the noise level of the continuum image, which behaves almost $\propto N^{-1/2}$ as well. 
In figure \ref{fig:skynoise} (Bottom), the noise level of the sky stack as a function of radius is shown for the $N_\mathrm{stack}=700$ case. Due to the pixel spatial correlation, they do not perfectly decrease as (the number of pixels in each annulus)$^{-0.5}$. Stacking methods again do not change the result.

At the same time, the average values of the sum of the counts in each annulus were measured. Due to systematic errors and sky residuals, the average sky counts are not exactly equal to zero. To correct this effect, we subtract the average sky value when we derive the radial profile. Typical sky value of the Ly$\alpha$ and continuum images are $\sim-5\times10^{-21}~\mathrm{erg~s^{-1}~cm^{-2}~arcsec^{-2}}$ and $\sim1.2\times10^{-32}~\mathrm{erg~s^{-1}~cm^{-2}~Hz^{-1}~arcsec^{-2}}$. 

\begin{figure}
\includegraphics[width=\columnwidth]{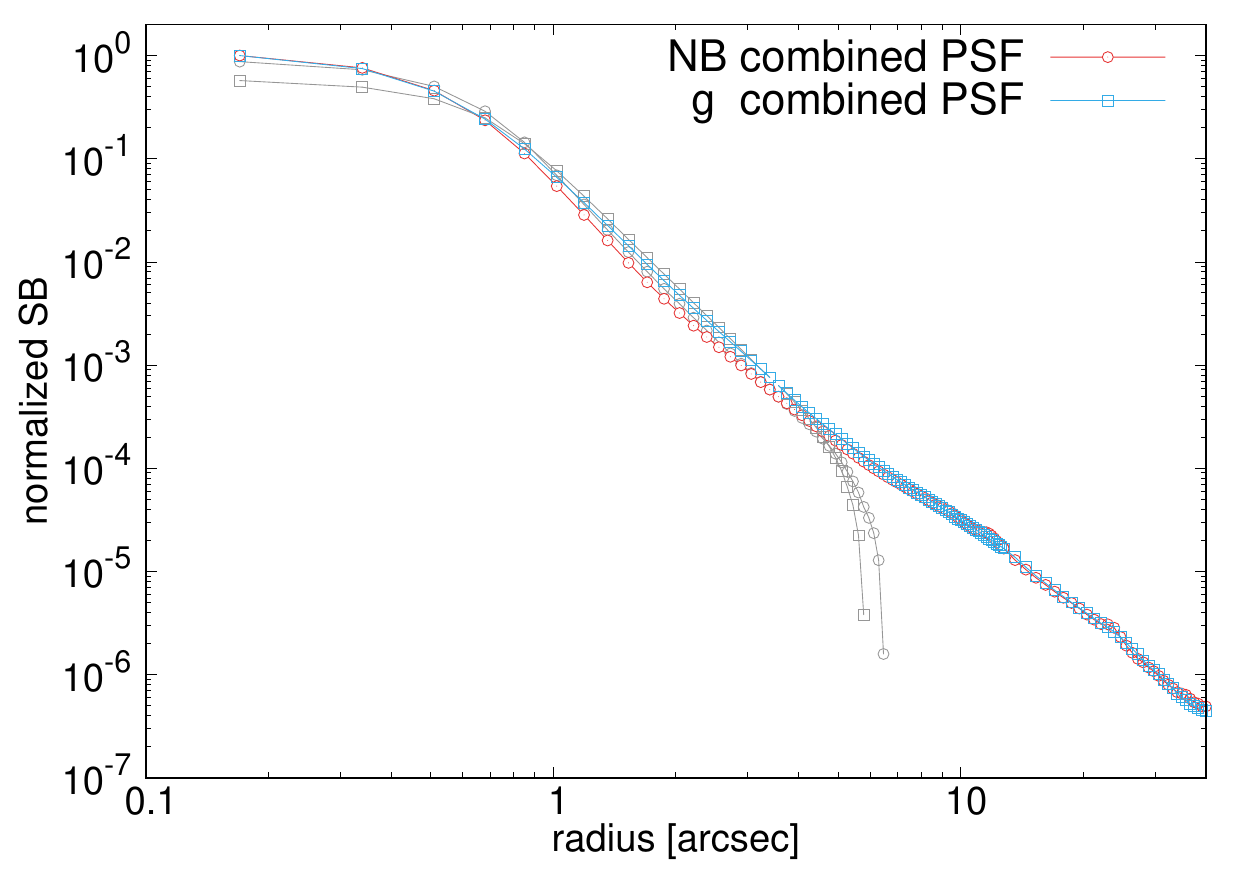}
  \caption{Normalized PSFs of the NB (red circles) and g (blue squares) band images. The inner and outer PSFs are connected at $r=3.4$ arcsec (20 pixels). Gray curves are the extrapolated inner and outer PSFs beyond the junction radius.  
\label{fig:3psf}}
\end{figure}

Since the Ly$\alpha$ image was created by subtracting the g-band image from the NB image, any difference between the PSFs of the images could produce spurious patterns around sources in the Ly$\alpha$ image. Even if the simple Gaussian smoothing done in Section \ref{sec:data} can match the FWHM of stellar sources, it cannot exactly match the shape of the PSFs of the two images. Moreover, the shape of the PSF at a large radius may introduce additional errors. 
To examine the detailed shapes of the PSF in the two images, we first select bright unsaturated sources from a source catalog using the SExtractor output CLASS\_STAR, which is a parameter characterizing the stellarity of sources. CLASS\_STAR is 1 if an object is a point source and drops to 0 if extended. Here we use following criteria: CLASS\_STAR$>0.95$ and $18<\mathrm{g}<22$. In total, 3980 sources are stacked to determine the central part of the PSFs in the images\footnote{Initially we divide this sample into two; one for sources distributed in the inner part of the field and the other for the outer part. The profiles of the stacked image of the two subsamples are almost identical. Thus we conclude that variation of the PSF within the field is minor and ignore the effect in the following analyses.}. 
To determine the much fainter outer part of the PSFs, we extracted stars with magnitude $13<\mathrm{g_{SDSS}}<15$ from the SDSS DR14 catalog \citep{Abolfathi2018}. After excluding stars with bright nearby companions and/or obviously extended sources when seen with our deep images, images of 113 bright stars are stacked. Since point sources in this magnitude range start to saturate, the PSF of the brighter sources is connected at $r=20$ pixels or 3.4 arcsec with that of fainter sources, following a method described in \citet{Infante-Sainz2019}. Derived PSFs from 0.17 arcsec to 40 arcsec are shown in Figure \ref{fig:3psf}.
The PSF of the NB image is slightly smaller than that of the g band at $r=$ a few arcsec. 
The PSFs of both bands beyond several arcsec agree very well. They are not Gaussian-like and have power-law tails with a slope of $\sim -2.8$.

\begin{figure}
\includegraphics[angle=270,width=\columnwidth]{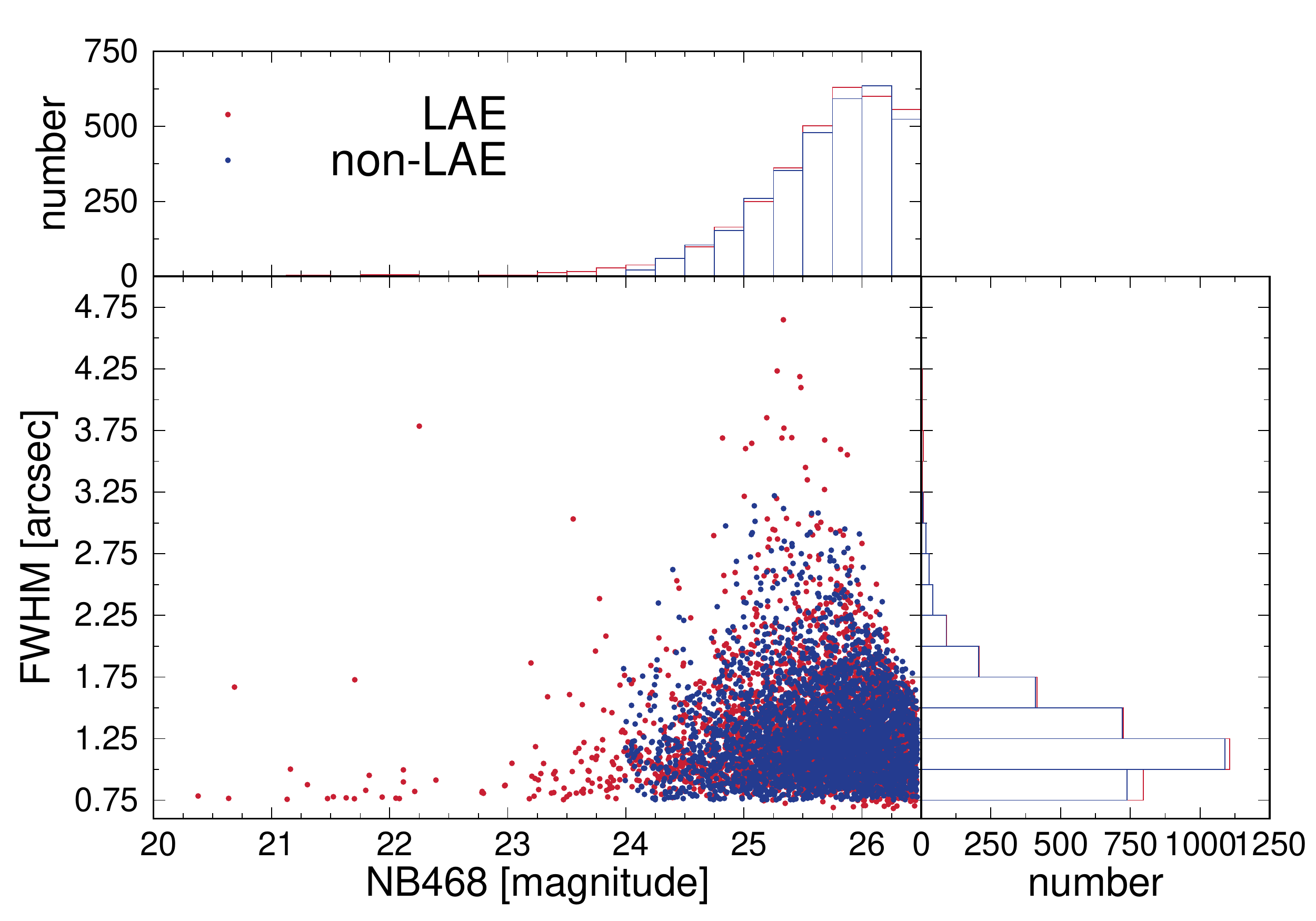}
  \caption{
  FWHM VS NB468 plot of LAE (red) and non-LAE (blue). Top and right panels show histograms of NB468 and FWHM, respectively. Their distribution is almost identical.
\label{fig:3nonlaehist}}
\end{figure}

To check whether the slight difference between the broadband and NB PSFs affects our surface brightness measurement, we created a stacked ``non-LAE'' image, following a method described in \citet{Momose2014}. ``Non-LAE'' sources are defined as objects not selected as LAEs which have almost the same distribution in the FWHM$_\mathrm{NB468}$ vs. NB468 magnitude plane as the real LAEs (Figure \ref{fig:3nonlaehist}). Since the majority of LAEs are distributed in the range $0.75\mathrm{~arcsec}<\mathrm{FWHM}<3.25$ arcsec and $24<\mathrm{NB468}<26.5$, we select non-LAE sources from this range for stacking. 
Any signal detected in the stacked Ly$\alpha$ image of non-LAEs can be used to estimate the effect not only of the PSF difference but also of other unknown systematics such as errors associated with flat-fielding and sky subtraction as discussed in \citet{Feldmeier2013}.

\section{Results\label{sec:result}}
\subsection{Stacked profiles and effects of systematics\label{sec:4systematics}}
\begin{figure}
\plotone{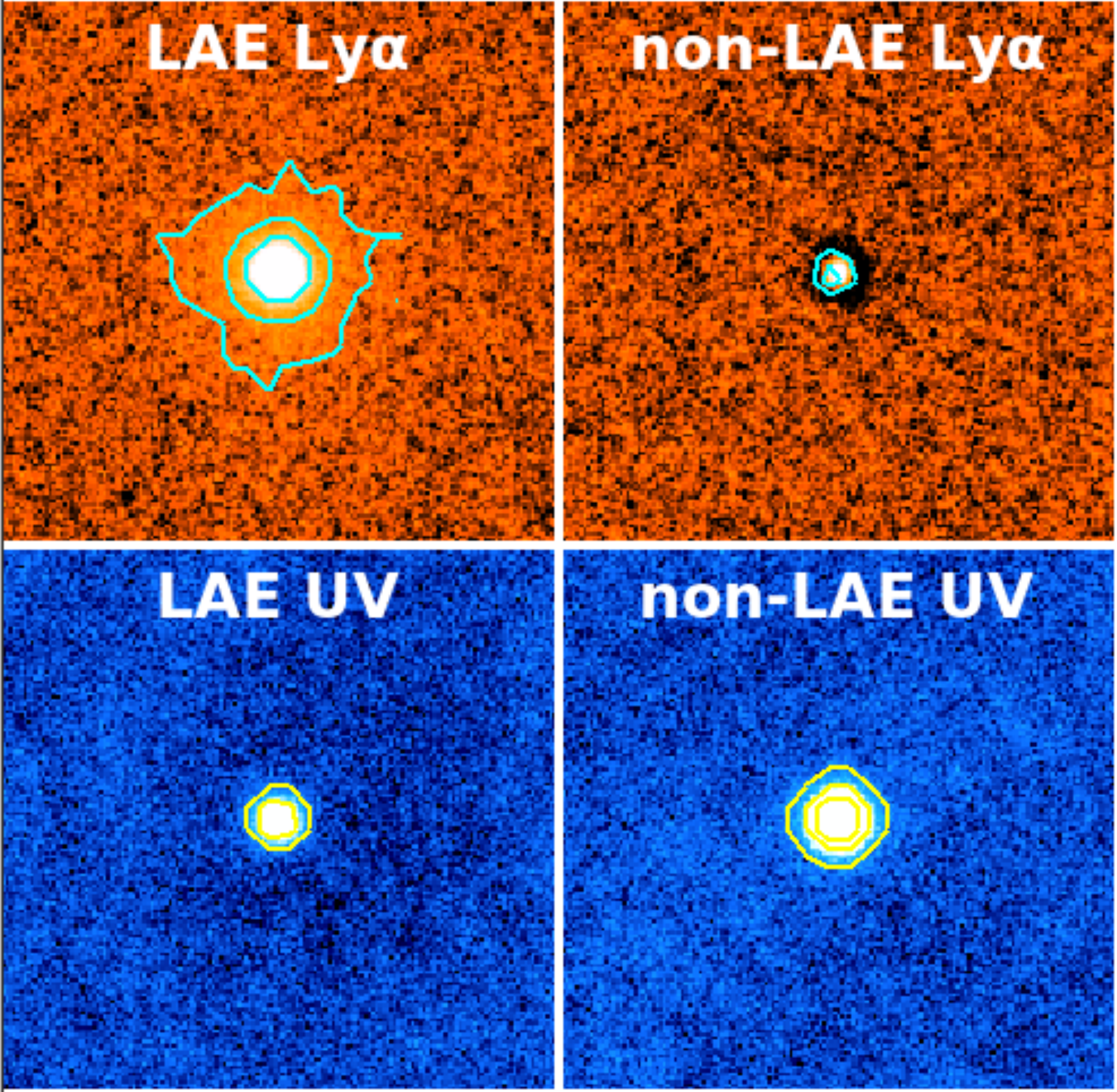}
  \caption{
  Stacked Ly$\alpha$ (top) and continuum (bottom) image of all LAEs (left) and non-LAEs (right).
  The size of each image is $\sim25\arcsec\times25\arcsec$ or $\sim200\times200$ pkpc at $z=2.84$.
  Contours correspond to $3\times10^{-18}, 1\times10^{-18}, 1\times10^{-19} \mathrm{~erg~s^{-1}~cm^{-2}~arcsec^{-2}}$ in the Ly$\alpha$ images and $3\times10^{-31}, 1\times10^{-31}, 3\times10^{-32} \mathrm{~erg~s^{-1}~cm^{-2}~Hz^{-1}~arcsec^{-2}}$ in contiuum images.  
\label{fig:3Lyaimageall}}
\end{figure}

\begin{figure}
\includegraphics[angle=270,width=\columnwidth]{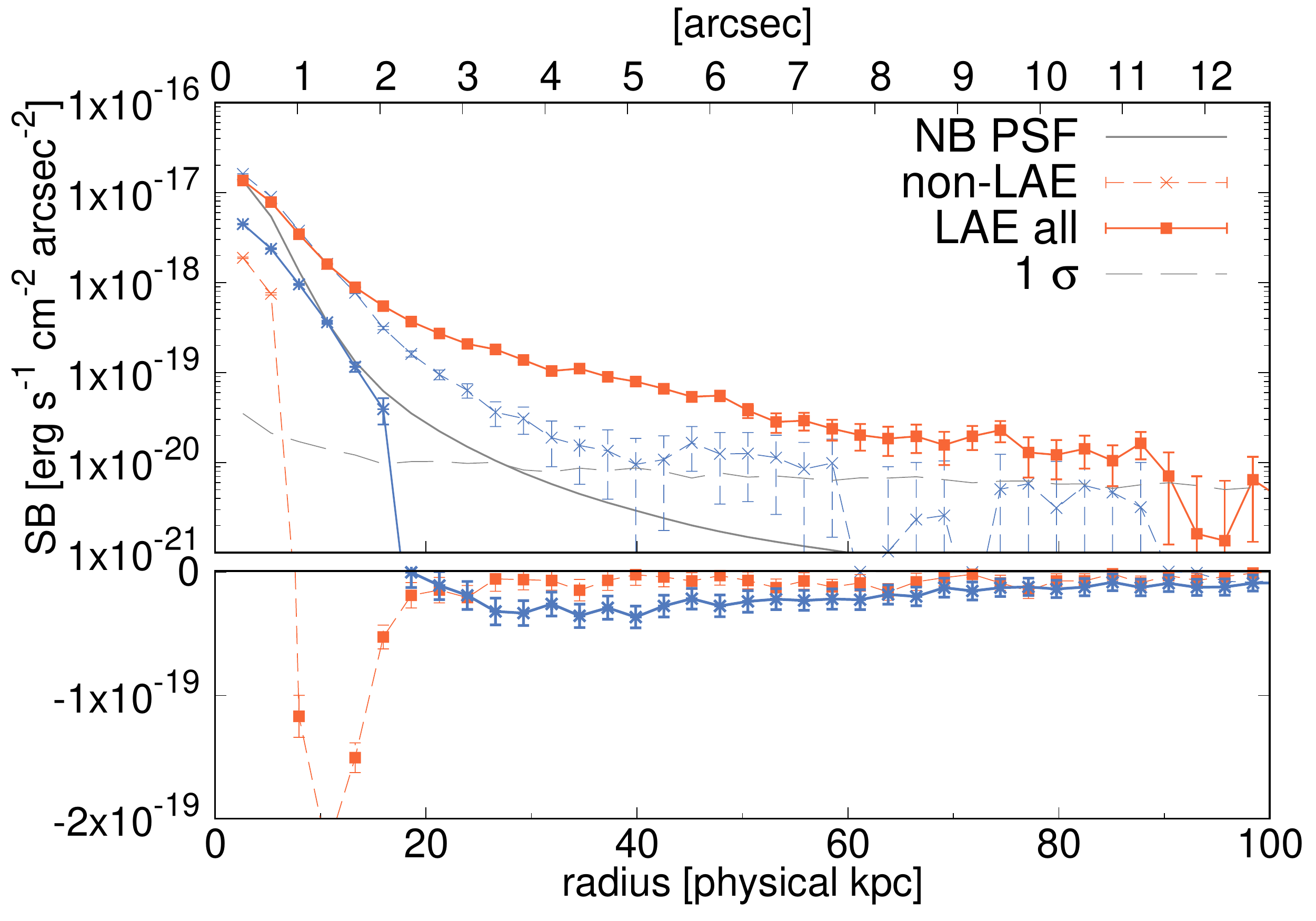}
  \caption{
  Radial SB profile of LAEs (solid) and non-LAEs (dashed) in the Ly$\alpha$ (orange) and continuum (blue curve) images are shown. Gray dashed curve shows 1$\sigma$ noise level. On the bottom side, SB profiles below $1\times10^{-21}\mathrm{~erg~s^{-1}~cm^{-2}~arcsec^{-2}}$ level are shown in linear scale. The normalized PSF of the NB image is shown with the gray solid curve. 
\label{fig:3Lyaall}}
\end{figure}

Figure \ref{fig:3Lyaimageall} shows the median-stacked Ly$\alpha$ and continuum images of all LAEs and non-LAEs without sigma clipping. The SB profiles of them are shown in Figure \ref{fig:3Lyaall}. We confirmed that the profiles do not depend on the stacking methods (except for average stacking without sigma clipping; see Figure \ref{fig:skynoise}); they show $>1\sigma$ deviation only at very low-S/N regime near $r\sim100$ pkpc. Hereafter, we present the results for median stacking without sigma clipping. 
The non-LAE has a negative ring-like structure around the center. This probably arises from the slight differences in their colors and PSFs of g-band and NB images. Still, the absolute value of the Ly$\alpha$ SB profile of the non-LAE is about an order of magnitude smaller than that of LAEs in Figure \ref{fig:3Lyaall}. Beyond 2 arcsec, the SB profile of the non-LAE stack is almost consistent with the sky value and thus we conclude the effect of the PSF difference is negligible, in particular at large radii of $r>2$ arcsec of most interest to the present work. 
The PSF, which is shown with the gray curve in Figure \ref{fig:3Lyaall}, drops much more rapidly than the Ly$\alpha$ profile of LAEs. From the above arguments, we conclude that LAHs around LAEs at $z=2.84$ are robustly detected down to $\sim1\times10^{-20}\mathrm{~erg~s^{-1}~cm^{-2}~arcsec^{-2}}$ and the effects of systematic errors cannot have a significant impact on the derived Ly$\alpha$ SB profiles out to $\sim100$ pkpc.

On the other hand, the UV SB profile of LAEs seems to be negative beyond 20 pkpc. 
Similar patterns can be also seen in some previous studies performing stacking analyses \citep{Matsuda2012,Momose2016} but the exact reasons were never identified in the literature. 
When estimating the sky value with SExtractor, pixels with counts above some threshold are masked. 
However, since our LAEs are selected with the NB image as a detection band, some LAEs in our sample are too faint in the UV continuum image to be masked. 
In addition, when creating the continuum image, the Ly$\alpha$ contribution is subtracted even if the source is not detected or significantly affected by the sky noise in the g-band image for UV-faint sources. 
These effects cause oversubtraction in the continuum image of UV-faint LAEs, affecting the UV SB profiles of subsamples that contain UV-faint LAEs. In our sample, there are 430 (1391) LAEs with $<2\sigma$ ($<5\sigma$) detection in the g-band image, respectively. The numbers of LAEs with $<2\sigma$ detection in all subsamples are also shown in Table \ref{tab:halostack}.
Thus, UV SB profiles of subsamples that contain many UV non-detected LAEs should be interpreted with caution
\footnote{It is also possible that a color term difference within the subsamples affects our SB measurement, although a correlation between $\beta$ and UV absolute magnitude $\mathrm{M_{UV}}$ of LAEs is, though still debated, weak \citep[$\mathrm{d}\beta/\mathrm{dM_{UV}}\sim-0.1$--0.0][]{Hathi2016, Hashimoto2017}}.

\begin{figure}
\includegraphics[width=\columnwidth]{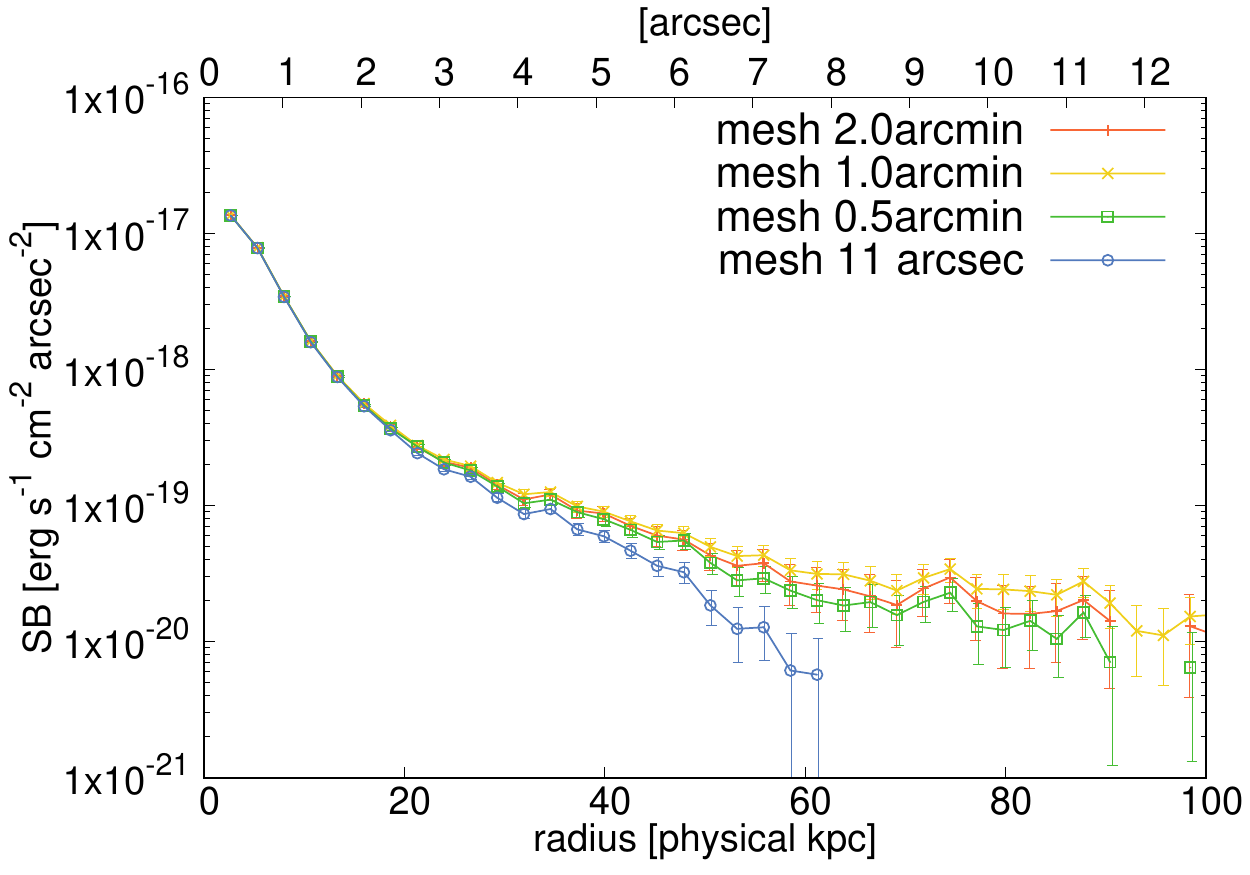}
  \caption{
  Radial Ly$\alpha$ SB profiles of all LAEs with different sky mesh size. Red, yellow, green, and blue curves respectively indicate profiles derived with sky mesh sizes of 2 arcmin, 1 arcmin, 0.5 arcmin, and 11 arcsec.
\label{fig:3Lyaskymesh}}
\end{figure}

To check whether or not the mesh size for sky estimation matters, we derived a stacked Ly$\alpha$ profile of all LAEs with sky mesh sizes different from 30 arcsec. A larger mesh size enables us to probe possible large-scale emission around LAEs, at the same time increasing errors due to residual non-astrophysical signals (which stem from e.g. halos around bright stars). A smaller mesh size may lead to oversubtraction of the real signal while reducing the errors described above.
To find a better compromise, we tested sky mesh sizes of 1 arcmin, 2 arcmin, and 11 arcsec (64 pixels, the default mesh size used in LAE selection in Section \ref{sec:data}). The 1$\sigma$ errors and residual sky emission to be subtracted were derived in the same way as described in Section \ref{sec:3anaunc}.
In Figure \ref{fig:3Lyaskymesh}, we showed the results of this test. Except for the case of 11 arcsec (blue curve), the derived Ly$\alpha$ SB profiles are all consistent with each other within uncertainty. 
Also, no systematic trend is evident with increasing mesh size in the slight offset seen in the outer part. This suggests the effect of oversubtraction of diffuse Ly$\alpha$ emission is minor at this sensitivity, on this scale. 
However, as the larger sky mesh sizes are utilized, residual artificial emission around bright stars in the Ly$\alpha$ images becomes clearer as well. 
On the other hand, a mesh size of 64 pixels $=11$ arcsec $=85$ pkpc clearly oversubtract halo emission. Considering that LAHs are detected out to $\sim100$ pkpc, a mesh size of 85 pkpc, which is comparable to the extent of LAHs, should not be used. 
The same trend is seen in the continuum image as well.
Thus we decided to use 0.5 arcmin $=$ 30 arcsec sky mesh.

\begin{figure*}
\includegraphics[width=\textwidth]{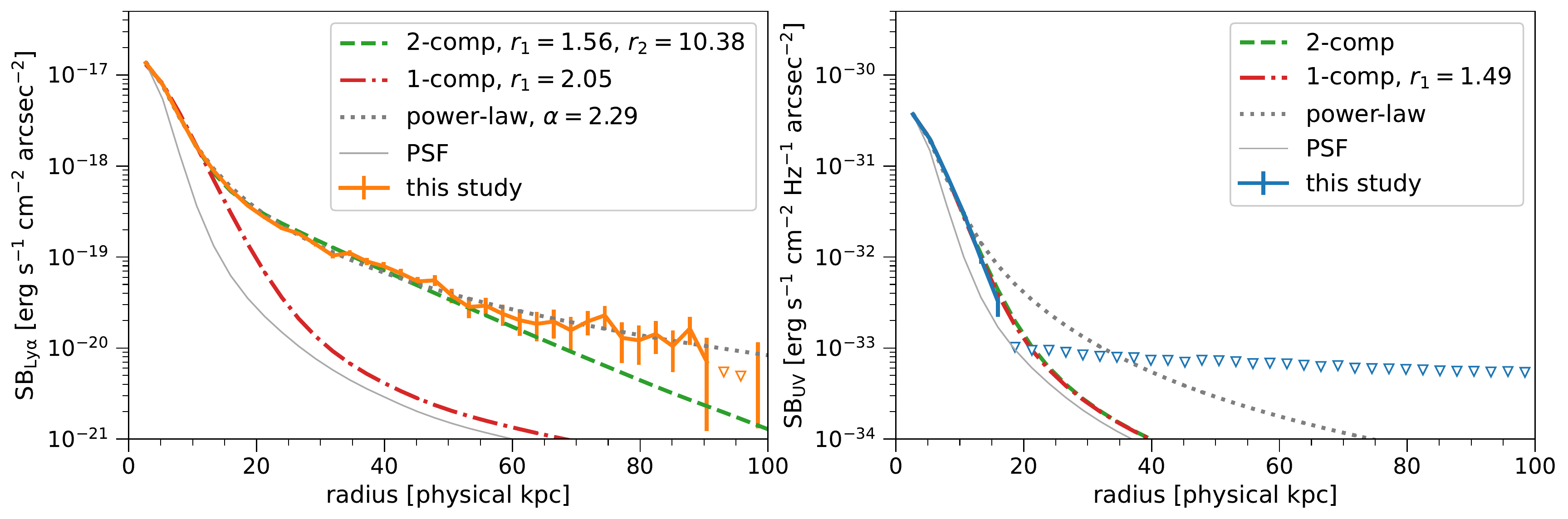}
  \caption{
  Radial Ly$\alpha$ (left) and UV (right) SB profiles of all LAEs are shown with solid orange/blue curves with errorbars. Dashed, dot-dashed, and dotted curves respectively show the result of fitting with two- and one-component exponential functions and a power-law function. The normalized PSF of the NB image is shown with the thin gray curve. 
  Downward triangles show 1$\sigma$ error levels after residual sky subtraction.
}
\label{fig:3Lyauvwfit}
\end{figure*}

To quantify the extent of SB profiles, we performed fits to both UV and Ly$\alpha$ SB profiles using the following exponential function(s) and power-law function:
\begin{eqnarray}
    \mathrm{PSF} * \left[ C_1 \exp{\left(-\frac{r}{r_1}\right)} + C_2 \exp{\left(-\frac{r}{r_2}\right)} \right] \label{eq:exp}\\
    \mathrm{PSF} * C_3 r^{-\alpha}, \label{eq:pow}
\end{eqnarray}
where ``$\mathrm{PSF}*$'' means convolution with the measured PSF of NB468.
While exponential functions have commonly been used in previous observational work, a power-law function is motivated by an analytical model by \citet{Kakiichi2018}.
$C_2$ is set to zero for 1-component exponential fitting and let $r_1<r_2$ if otherwise, thus $r_1$ is scale-length for the core component and $r_2$ is for the halo component. 
Unlike most previous work, we do not assume that the scale-lengths for the core component of Ly$\alpha$ and UV SB profile are the same, and they were fitted separately. 
The result shown in Figure \ref{fig:3Lyauvwfit} clearly demonstrates the need for non-zero $C_2$ or a power-law function to fit the Ly$\alpha$ SB profile, while a single exponential function will do for the UV SB profile fitting. The 2-component exponential fit to the UV SB profile does not converge, and the power-law fit clearly deviates from the observed UV profile.

\subsection{Subsamples}\label{sec:ressub}
\begin{figure*}
\plotone{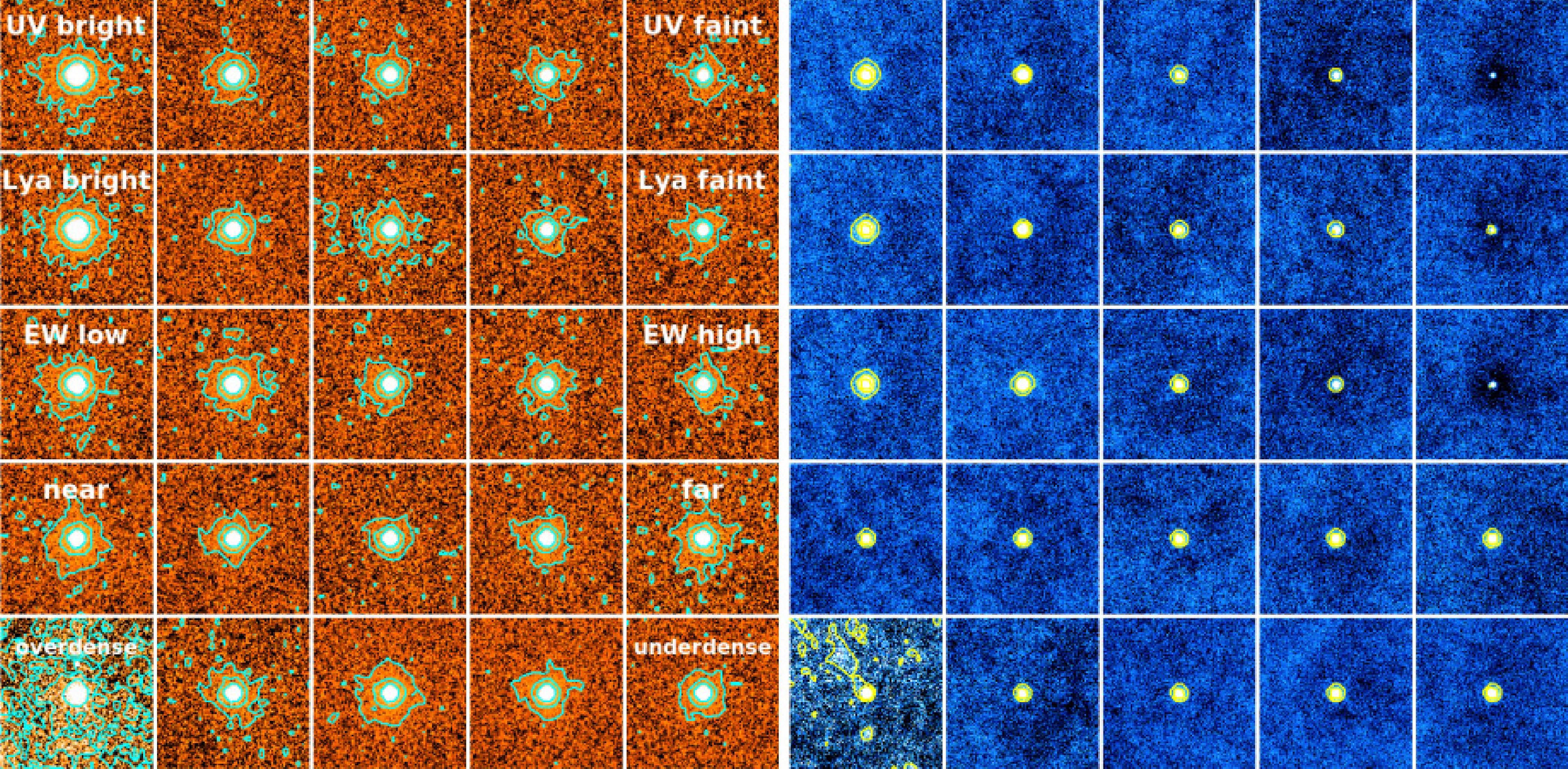}
\caption{
  Stacked Ly$\alpha$ (left, orange) and continuum (right, blue) images of different subsamples in the asinh color stretch. From top to bottom, we show stacked images of UV magnitude, $L_\mathrm{Ly\alpha}$, EW$_\mathrm{0,Ly\alpha}$, distance from the HLQSO, and environment subsamples. From left to right, median [UV magnitude, $L_\mathrm{Ly\alpha}$, EW$_\mathrm{0,Ly\alpha}$, distance from the HLQSO, $\delta$] of each subsample respectively become [fainter, fainter, larger, larger, smaller]. The size of each image is $\sim200 \times 200$ pkpc. 
  Contours correspond to $3\times10^{-18}, 1\times10^{-18}, 1\times10^{-19} \mathrm{~erg~s^{-1}~cm^{-2}~arcsec^{-2}}$ in the Ly$\alpha$ images and $3\times10^{-31}, 1\times10^{-31}, 3\times10^{-32} \mathrm{~erg~s^{-1}~cm^{-2}~Hz^{-1}~arcsec^{-2}}$ in continuum images. . 
}
\label{fig:3Lyaimage1}
\end{figure*}

\begin{figure*}
\includegraphics[width=\textwidth]{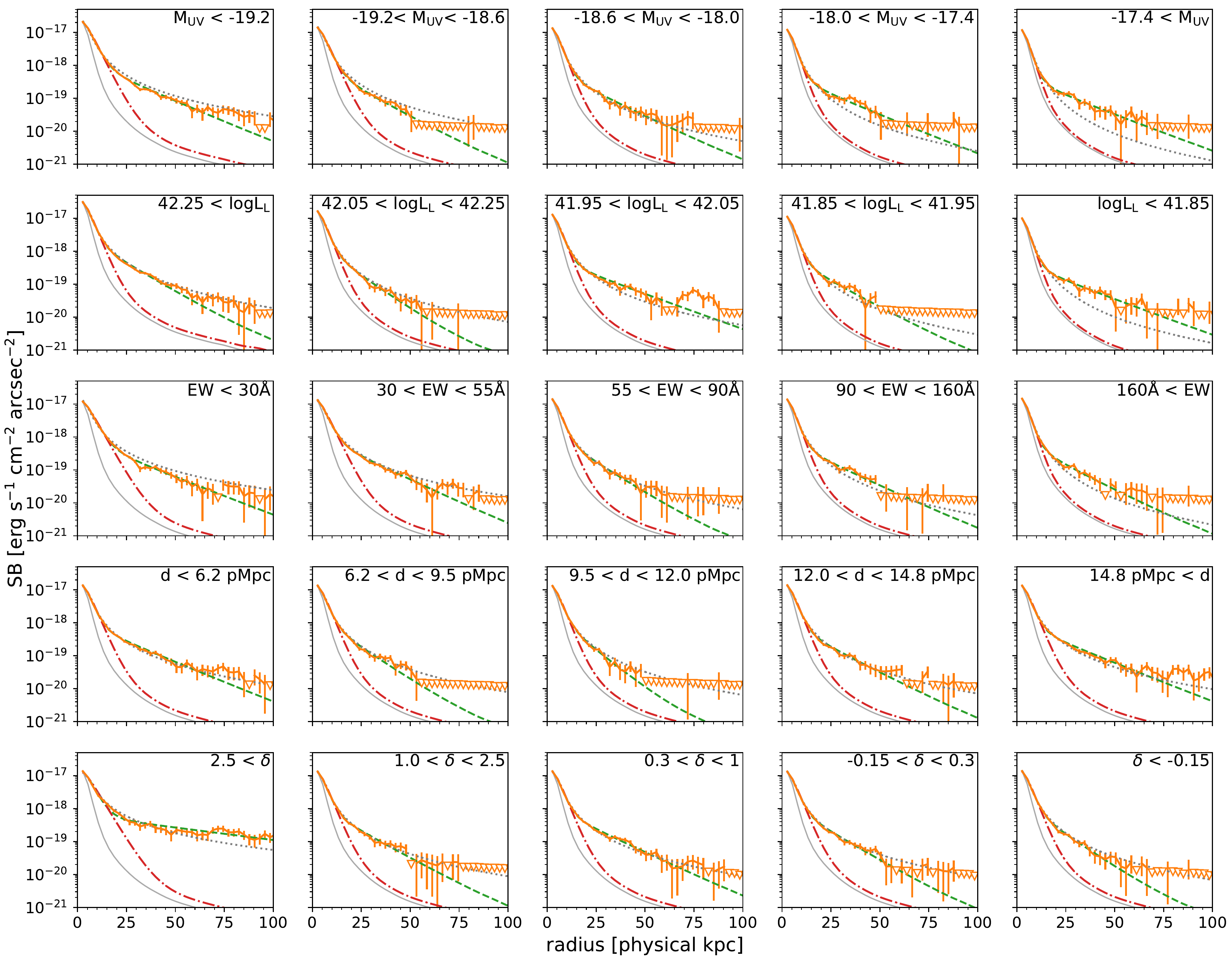}
  \caption{
  Radial Ly$\alpha$ SB profiles of all subsamples (specified in Table \ref{tab:halostack}, solid orange curve with errorbars) with fitting curves. Green dashed, red dot-dashed, and gray dotted curves show the result of fitting with two- and one-component exponential functions and a power-law function. Downward triangles show 1$\sigma$ error levels after residual sky subtraction. 
  Thin gray curve shows the normalized PSF shape.
\label{fig:3Lyawfit}}
\end{figure*}

\begin{figure*}
\includegraphics[width=\textwidth]{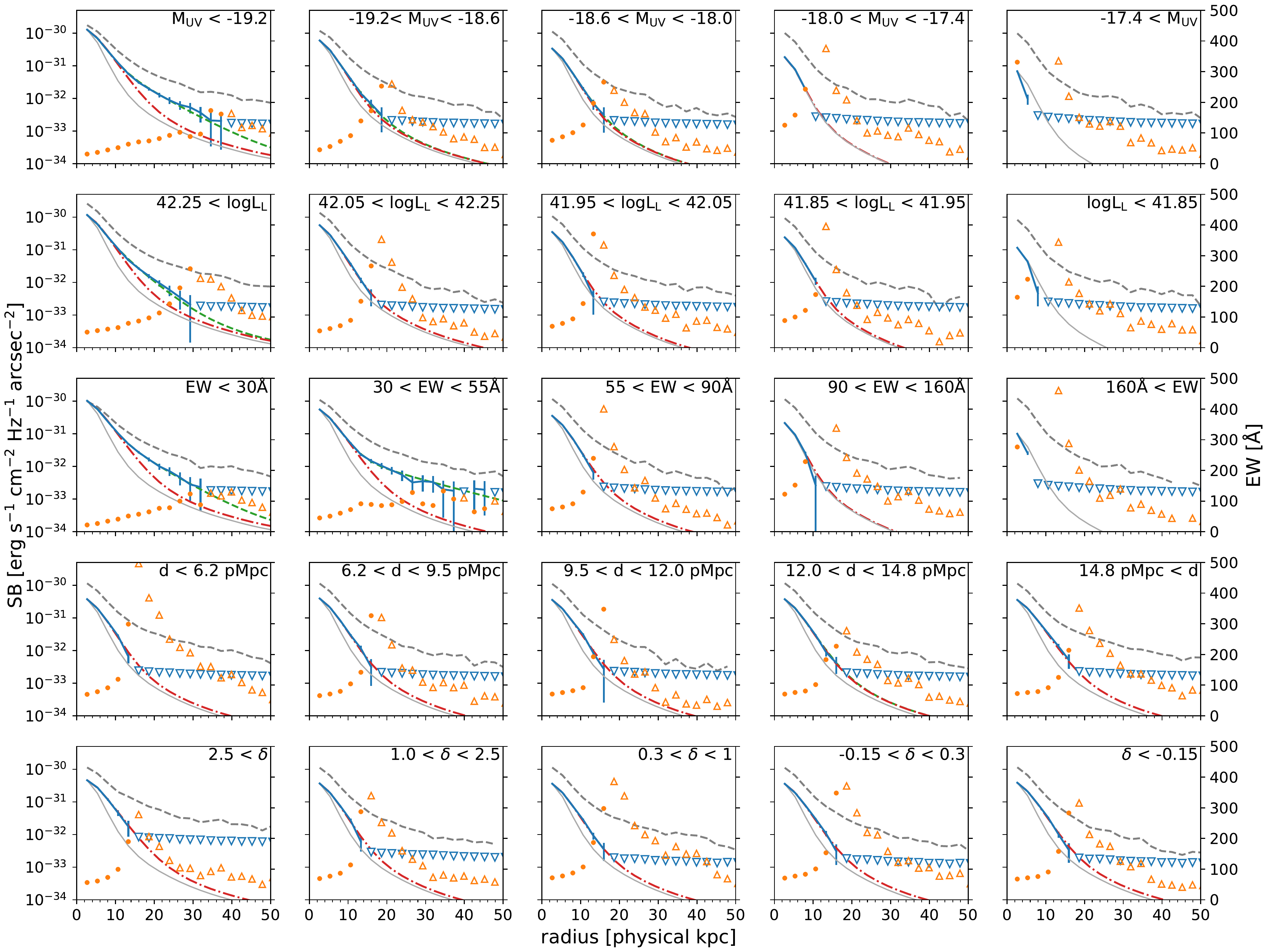}
  \caption{
  Radial UV SB profiles of all subsamples (specified in Table \ref{tab:halostack}, solid blue curve with errorbars) with fitting curves. Green dashed and red dot-dashed curves show the result of fitting with two- and one-component exponential functions. 
  The gray dashed curve is Ly$\alpha$ SB profiles converted from $F_\mathrm{Ly\alpha}$ units to $f_\nu$ units by simply dividing by the FWHM of the NB filter. Orange dot indicates rest-frame Ly$\alpha$ equivalent width in \AA calculated in each annulus, with its value on the right axis. Downward and upward triangles show 1$\sigma$ limits of UV emission and equivalent width (some datapoints in the rightmost panels are above the upper boundary) after residual sky subtraction. 
  Thin gray curve shows the normalized PSF shape.
  Note the difference in the range of the x-axis from that of Figure \ref{fig:3Lyawfit}. 
\label{fig:3uvwfit}}
\end{figure*}

The stacked Ly$\alpha$ and UV continuum images of all subsamples are shown in Figure \ref{fig:3Lyaimage1}; their corresponding Ly$\alpha$ and UV continuum SB profiles are shown in Figures \ref{fig:3Lyawfit} and \ref{fig:3uvwfit}, respectively with fitting curves.  
Rest-frame Ly$\alpha$ equivalent widths calculated in each annulus are also plotted with orange dots in Figure \ref{fig:3uvwfit}. 
The resulting fit parameters for the Ly$\alpha$ SB profiles are given in Table \ref{tab:halostack}, and those of UV SB profiles in Table \ref{tab:resuv1}.
In the right panel of Figure \ref{fig:3Lyaimage1}, 
the effect of oversubtraction discussed in Section \ref{sec:4systematics} is clearly manifested by the black ring-like structures around the central emission in the stacked UV continuum images of the UV/Ly$\alpha$ faintest and highest-EW subsamples (rightmost three panels from top to middle). 
Again, the UV SB (and the EW) profiles of UV-faint subsamples should be interpreted with caution.

We detected Ly$\alpha$ emission more extended than UV (stellar) emission in all subsamples (Figure \ref{fig:3Lyawfit}), while most UV SB profiles can be fitted with the one-component exponential function (Figure \ref{fig:3uvwfit}). 
However, UV SB profiles of the UV/Ly$\alpha$ brightest and lowest-EW subsamples clearly require two-component exponential functions and indeed can be fitted remarkably well. To our knowledge, this is the first detection of ``UV halos'' in high-redshift LAEs. 
We confirmed that this detection is robust against the choice of used stacking methods, the sky mesh size (as long as it is not too small), and the masking threshold. In addition, these subsamples are securely detected in the g-band and thus the effect of oversubtraction (Section \ref{sec:4systematics}) should be minor. 
For subsamples for which two-component exponential fitting is well converged, we show the resultant fitting parameters in Table \ref{tab:resuv2}. 
Figures \ref{fig:Lyaprofs1} and \ref{fig:Lyaprofs2} in Appendix \ref{sec:lyasbsep} show results in a different manner for a clearer comparison within each photometric property. In Figure \ref{fig:Lyaprofs1}, clear systematic differences can be seen in bins of UV, $L_\mathrm{Ly\alpha}$, and EW$_\mathrm{0,Ly\alpha}$ in a way that UV/Ly$\alpha$-bright LAEs and low EW$_\mathrm{0,Ly\alpha}$ LAEs have larger LAHs. 
We see hints of UV halos also in the second and third UV brightest subsamples in the upper right panel of Figure \ref{fig:Lyaprofs1}.
On the other hand, the difference of profiles for the projected distance and local environment subsamples is not obvious except for the protocluster subsample (those with $\delta>2.5$). 
There is no significant difference in UV SB profiles in both cases except for the protocluster subsample, which contains more UV bright galaxies as seen in Figure \ref{fig:cumdis0}.

\begin{figure*}
\includegraphics[width=\textwidth]{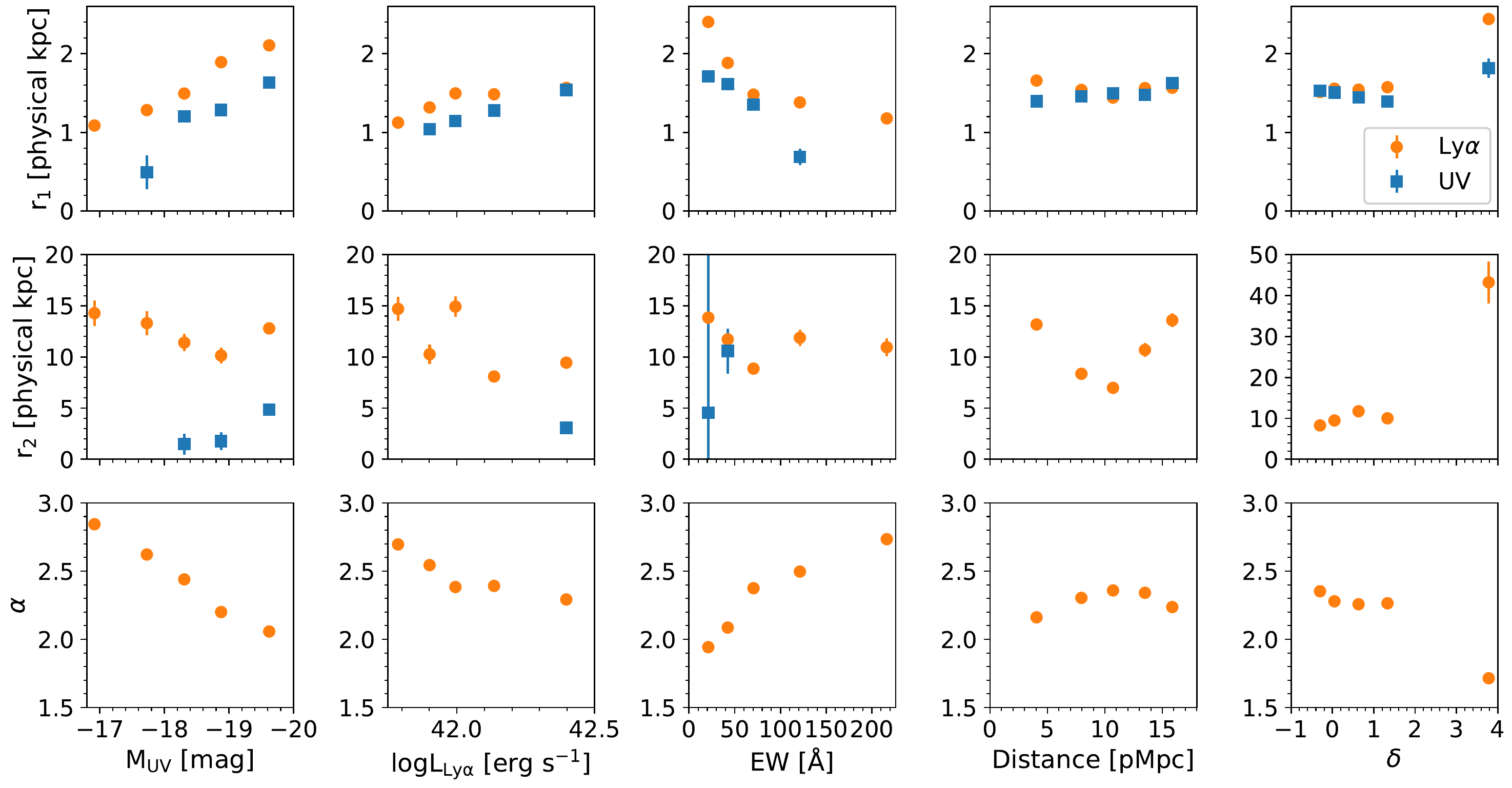}
  \caption{
  Resulting scale-lengths and power-law index vs. median of each subsample. From left to right, we show the results for UV magnitude, Ly$\alpha$ luminosity, Ly$\alpha$ equivalent width, distance from the HLQSO, and environment subsamples. $r_\mathrm{1,UV}$ are those obtained from a one-component fit, while $r_\mathrm{1,Ly\alpha}$ are from a two-component fit.
\label{fig:3scalelp}}
\end{figure*}

\begin{figure*}
\plotone{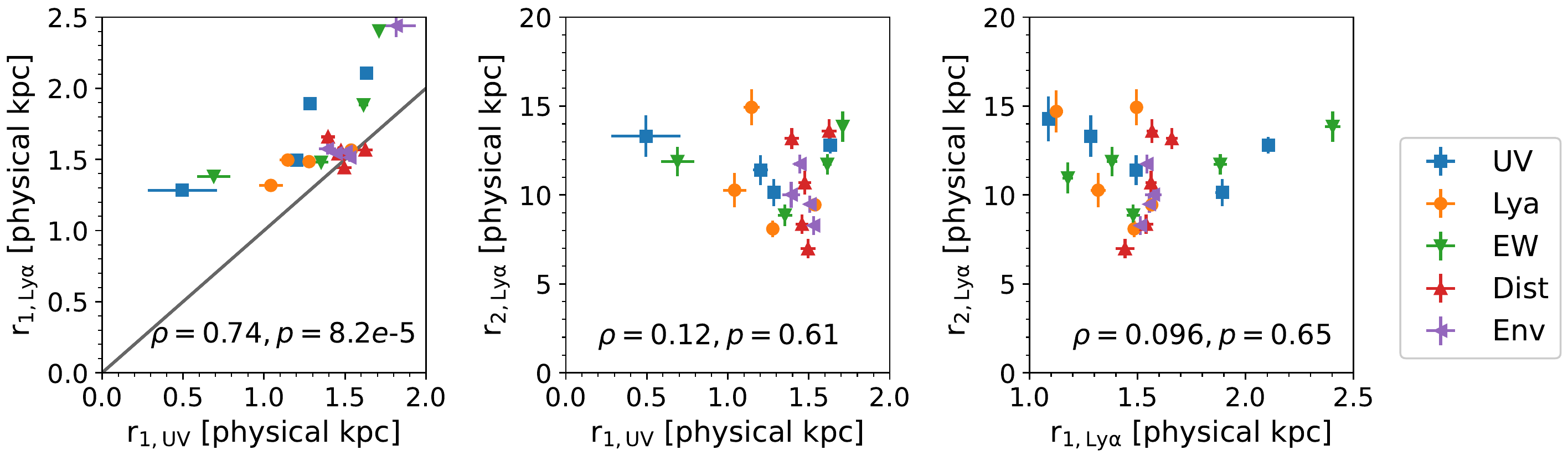}
  \caption{
  Comparison between scale-length for the UV first component $r_\mathrm{1,UV}$ and the Ly$\alpha$ first and second component $r_\mathrm{1,Ly\alpha}, r_\mathrm{2,Ly\alpha}$.
  $r_\mathrm{1,UV}$ are those obtained from a one-component fit, while $r_\mathrm{1,Ly\alpha}$ are from a two-component fit.
  In each panel, we list Spearman's rank correlation coefficient $\rho$ and p-value $p$. 
  The diagonal line in the left panel shows the 1:1 relation. 
\label{fig:3scalelc}}
\end{figure*}

The scale-lengths and power-law index of fitting functions can be used for more quantitative discussion. 
In Figure \ref{fig:3scalelp}, we show the results of fitting for each photometric property. 
$r_\mathrm{1,UV}$ plotted in Figures \ref{fig:3scalelp} and \ref{fig:3scalelc} are those obtained from a one-component fit (including subsamples with UV-halo), while $r_\mathrm{1,Ly\alpha}$ are from a two-component fit.
In the left three columns of the top row, both $r_\mathrm{1,Ly\alpha}$ and $r_\mathrm{1,UV}$ show nearly monotonic behavior. While the power-law index $\alpha$ also shows consistent behavior, the power-law functions often deviate from the real data beyond a few tens of pkpc in Figure \ref{fig:3Lyawfit}. 
On the other hand, $r_\mathrm{2,Ly\alpha}$ behaves not that simply. This would result from both astrophysical and observational reasons as we discuss in Section \ref{sec:5dependence}.

Figure \ref{fig:3scalelc} compares scale-lengths of UV/Ly$\alpha$ core/halo components. 
In the left panel, the scale-lengths for the core components are compared. The centroids of UV continuum emission of LAEs are known to show some offset from those of Ly$\alpha$ emission which are defined as the image centers in this work. The vast majority have offset lower than $0.2$ arcsec \citep{Shibuya2014, Leclercq2017} which is comparable to HSC's pixel scale of 0.17 arcsec. This should not affect the measurement of $r_\mathrm{2,UV}$ but leads to an overestimation of $r_\mathrm{1,UV}$. In addition, some subsamples contain g-band nondetected sources (see Table \ref{tab:halostack}) and could be affected by oversubtraction. 
Although one should be aware of these potential issues, $r_\mathrm{1,Ly\alpha}$ seems to be almost always larger than $r_\mathrm{1,UV}$ and they correlate with Spearman's rank correlation coefficient of 0.74 and p-value of $8.2\times10^{-5}$. 
On the other hand, similarly to the trend seen in the panels in the middle row of Figure \ref{fig:3scalelp}, $r_\mathrm{2,Ly\alpha}$ do not have a clear correlation with $r_\mathrm{1,UV}$ nor $r_\mathrm{1,Ly\alpha}$ ($p$-value 0.61 and 0.65, respectively).

\begin{deluxetable}{ccc}
\tabletypesize{\scriptsize}
\tablecaption{The result of the fitting of one-component exponential functions to UV profiles.}
\tablehead{
\colhead{criteria} & \colhead{$C_\mathrm{1,UV}$} & \colhead{$r_\mathrm{1,UV}$} \\
\colhead{(1)} & \colhead{(2)} & \colhead{(3)}
}
\startdata
All & $ 8.146 \pm 0.072 $ & $ 1.486 \pm 0.019 $ \\ \hline 
$M_\mathrm{UV}<-19.2$           & $26.830 \pm 0.144$  & $1.632 \pm 0.013$  \\
$-19.2<M_\mathrm{UV}<-18.6$     & $13.896 \pm 0.178$  & $1.285 \pm 0.027$  \\
$-18.6 < M_\mathrm{UV}<-18.0$   & $8.049 \pm 0.179$  & $1.203 \pm 0.046$  \\
$-18.0<M_\mathrm{UV}<-17.4$     & $5.552 \pm 0.309$  & $0.495 \pm 0.214$  \\
$-17.4<M_\mathrm{UV}$           & --  & --  \\ \hline
$42.25<\log L_\mathrm{Ly\alpha} \mathrm{[erg~s^{-1}]}$        & $25.078 \pm 0.157$  & $1.538 \pm 0.014$  \\
$42.05<\log L_\mathrm{Ly\alpha}<42.25$  & $13.285 \pm 0.164$  & $1.278 \pm 0.026$  \\
$41.95<\log L_\mathrm{Ly\alpha}<42.05$  & $8.606 \pm 0.210$  & $1.147 \pm 0.050$  \\
$41.85<\log L_\mathrm{Ly\alpha}<41.95$  & $6.118 \pm 0.220$  & $1.042 \pm 0.073$  \\
$\log L_\mathrm{Ly\alpha}<41.85$        & -- & --  \\ \hline
$12 < \mathrm{EW_{0,Ly\alpha}}<30$ \AA  & $21.208 \pm 0.142$  & $1.710 \pm 0.016$  \\
$30 <\mathrm{EW_{0,Ly\alpha}}<55$ \AA   & $11.772 \pm 0.141$  & $1.614 \pm 0.028$  \\
$55 <\mathrm{EW_{0,Ly\alpha}}<90$ \AA   & $8.182 \pm 0.170$  & $1.353 \pm 0.045$  \\
$90 <\mathrm{EW_{0,Ly\alpha}}<160$ \AA  & $6.173 \pm 0.264$  & $0.690 \pm 0.102$  \\
160 \AA $<\mathrm{EW_{0,Ly\alpha}}$     & --  & --  \\ \hline
$d_\mathrm{Q} < 6.2$ pMpc               & $8.451 \pm 0.168$  & $1.395 \pm 0.043$  \\
$6.2 < d_\mathrm{Q} < 9.5$ pMpc         & $8.716 \pm 0.154$  & $1.458 \pm 0.039$  \\
$9.5 < d_\mathrm{Q} < 12.0$ pMpc        & $7.798 \pm 0.163$  & $1.495 \pm 0.047$  \\
$12.0 < d_\mathrm{Q} < 14.8$ pMpc       & $8.229 \pm 0.149$  & $1.475 \pm 0.040$  \\
$14.8 pMpc < d_\mathrm{Q}<16.9$ pMpc    & $7.512 \pm 0.148$  & $1.625 \pm 0.046$  \\ \hline
$2.5 < \delta$          & $9.423 \pm 0.471$  & $1.815 \pm 0.123$  \\
$1.0 < \delta < 2.5$    & $8.319 \pm 0.212$  & $1.391 \pm 0.054$  \\
$0.3 < \delta < 1.0$    & $8.050 \pm 0.143$  & $1.445 \pm 0.038$  \\
$-0.15 < \delta < 0.3$  & $7.828 \pm 0.129$  & $1.507 \pm 0.036$  \\
$-1.0 < \delta < -0.15$ & $8.431 \pm 0.133$  & $1.530 \pm 0.034$  \\
\enddata
\tablecomments{Column (1): criteria used to define subsamples, Column (2): $C_\mathrm{1,UV}$ in units of $10^{-31} \mathrm{~erg~s^{-1}~cm^{-2}~Hz^{-1}~arcsec^{-2}}$, Column (3): $r_\mathrm{1,UV}$ in units of physical kpc. 
The uncertainties of fitting parameters include fitting errors only.
}
\label{tab:resuv1}
\end{deluxetable}

\begin{deluxetable}{ccc}
\tabletypesize{\scriptsize}
\tablecaption{The result of the fitting of two-component exponential functions to UV profiles.}
\tablehead{
\colhead{criteria} & \colhead{$C_\mathrm{1,UV}$, $C_\mathrm{2,UV}$} & \colhead{$r_\mathrm{1,UV}$, $r_\mathrm{2,UV}$} \\
\colhead{(1)} & \colhead{(2)} & \colhead{(3)}
}
\startdata
$M_\mathrm{UV}<-19.2$ & $26.730 \pm 0.389$  & $1.280 \pm 0.048$  \\ 
& $1.969 \pm 0.495$  & $4.845 \pm 0.532$  \\ 
$42.25<\log L_\mathrm{Ly\alpha} \mathrm{[erg~s^{-1}]}$ & $22.466 \pm 1.150$  & $0.874 \pm 0.110$  \\ 
& $5.483 \pm 0.495$  & $3.072 \pm 0.289$  \\ 
$12 < \mathrm{EW_{0,Ly\alpha}}<30$ \AA & $20.804 \pm 0.531$  & $1.343 \pm 0.071$  \\ 
& $1.818 \pm 0.660$  & $4.570 \pm 0.679$  \\ 
$30 <\mathrm{EW_{0,Ly\alpha}}<55$ \AA & $12.102 \pm 0.169$  & $1.407 \pm 0.050$  \\ 
& $0.250 \pm 0.080$  & $10.586 \pm 2.201$  \\ 
\enddata
\tablecomments{Column (1): criteria used to define subsamples, Column (2): First and second rows respectively show $C_\mathrm{1,UV}$ and $C_\mathrm{2,UV}$ in units of $10^{-31} \mathrm{~erg~s^{-1}~cm^{-2}~Hz^{-1}~arcsec^{-2}}$, Column (3): First and second rows respectively show $r_\mathrm{1,UV}$ and $r_\mathrm{2,UV}$ in units of physical kpc. 
The uncertainties of fitting parameters include fitting errors only.
}
\label{tab:resuv2}
\end{deluxetable}

\section{Discussion\label{sec:discussion}}

\subsection{Sources of Differences from Previous Observational Studies\label{sec:5prev}}
While we clearly detect LAHs more extended than UV continuum for all subsets of LAEs, some previous works report non-detections of such components \citep[e.g.,][]{Bond2010,Feldmeier2013}. We describe a number of possible reasons for such discrepancies between this study and others. 
The Ly$\alpha$ morphology cannot be properly captured by just comparing simple quantities such as their FWHMs or half-light radii without enough sensitivity or taking a very large aperture for total luminosity estimate \citep{Nilsson2009}. 
Detailed analyses on SB profiles are desirable but with enough sensitivity of (at least) $\sim10^{-19}\mathrm{~erg~s^{-1}~cm^{-2}~arcsec^{-2}}$, and even higher sensitivity is required for safer arguments beyond a mere detection. 
The scale-lengths of an exponential function(s) are most widely used in the literature for such analyses \citep[e.g.,][]{Steidel2011,Matsuda2012,Momose2014,Momose2016,Wisotzki2016,Leclercq2017,Xue2017}. 
However, resulting scale-lengths vary a lot depending not only on the data quality but also on the details of analyses: results depend on sample selection criteria, sky subtraction, masking, and stacking methods, the range of radius used for fitting, the radial binning size, fitting functions, whether to assume $r_\mathrm{1,UV}=r_\mathrm{1,Ly\alpha}$ or not, etc. 
First, we investigated the impact of the sensitivity using randomly chosen LAEs with a smaller sample size. We randomly took 100 LAEs and obtained $r_\mathrm{2,Ly\alpha}$ of stacked Ly$\alpha$ images, and repeated this process 1000 times. While the medians of obtained distribution of $r_\mathrm{2,Ly\alpha}$ did not show a systematic trend, the distribution had a 3.4 times larger standard deviation compared to that obtained for the 700 LAE case (see Section \ref{sec:5distance}), although this number may just represent diversity in our sample. Results from lower-sensitivity data could be thus more uncertain.
Secondly, most of the observed Ly$\alpha$ SB profiles are downwardly convex (Figure \ref{fig:3Lyawfit}) due to the flattening at $\sim15$ pkpc and thus the scale-length becomes smaller when the outer (inner) boundary of the fitting range is smaller (larger). 
Indeed, if we limit our fitting range up to $<30$ pkpc, obtained $r_\mathrm{2,Ly\alpha}$ is underestimated by 35\%. 
The outer boundary is also affected by the sensitivity; without deep data, one has no choice other than to set it to smaller values where the signal is detected. 
Thirdly, as we showed in Figure \ref{fig:3Lyaskymesh}, an insufficiently small mesh size for local sky background estimate leads to underestimation of the scale-length of LAHs, but in many cases the mesh size (or even a brief summary about sky subtraction) has not been presented in the text. 
As for fitting functions, a two-component fit can more robustly capture the shape of the Ly$\alpha$ SB profiles \citep[see Appendix C of][]{Xue2017}. 
However, there are few cases where such analyses have been present with enough sensitivity at $z\sim3$. 

For example, \citet{Xue2017} reported a halo scale-length of LAEs of $\sim5$--9 pkpc and found no evidence for environmental dependence based on NB surveys of two protoclusters, at $z=3.78$ and $z=2.66$. These observations are an order of magnitude shallower than the present study; \citet{Xue2017} used the Mayall 4m telescope for the $z=3.78$ data, and the Subaru telescope for the $z=2.66$ data but used an intermediate band filter (IA445 on Suprime-Cam, $\Delta\lambda=201$\AA) which has lower line sensitivity. 
The number of LAEs used to examine environmental dependence was at most $139$ (for an intermediate density sample). 
The criteria used to select LAEs picks up relatively high EW LAEs, with EW$_\mathrm{0,Ly\alpha}>50$\AA, in the $z=2.66$ protocluster field. Our study suggests that this could have biased the results toward smaller LAHs (Figure \ref{fig:Lyaprofs1}). 
Lower ionizing radiation field strength and/or lower abundance of cool gas in their protoclusters may also produce smaller LAHs (see Section \ref{sec:5distance} and \ref{sec:5environ}). 
As for observations with integral field spectrographs, \citet{Wisotzki2016,Leclercq2017} probed $r<30$ pkpc of LAEs at $z>3$ and obtained $r_\mathrm{2,Ly\alpha}$ of $\lesssim15$ pkpc with the majority with $r_\mathrm{2,Ly\alpha}<5$ pkpc. Analyses on an individual basis would suffer from greater noise and sample variance than ours.
On the other hand, \citet{Chen2021azim} probed 59 star-forming galaxies (including non-LAEs) at $z=2$--3 with sufficiently deep Keck/KCWI observational data and conducted stacking. They got $r_\mathrm{1,Ly\alpha}=3.71^{+0.06}_{-0.04}$ pkpc and $r_\mathrm{2,Ly\alpha}=15.6^{+0.5}_{-0.4}$ pkpc for their stacked Ly$\alpha$ SB profiles. Their sample is typically about an order of magnitude more massive and star-forming than our sample and this could explain the size difference. 
To summarize, comparing results obtained using inhomogeneous analyses in the literature without consistent reanalysis is difficult, and thus we will not attempt a detailed comparison here.

\subsection{Dependence of Scale-length on Galaxy Properties}\label{sec:5dependence}
In Figure \ref{fig:3scalelp}, UV/Ly$\alpha$-bright or low-EW LAEs tend to have larger $r_{1,Ly\alpha}$ and $r_{1,UV}$ than UV/Ly$\alpha$-faint or high-EW LAEs. This is qualitatively consistent with the HST-based results of \citet{Leclercq2017} although our results from a ground-based telescope tend to show larger values (0.1--1 pkpc vs $>1$ pkpc). Considering that UV/Ly$\alpha$ luminous and/or low-EW LAEs tend to be more massive, the trend is also consistent with the known trend between $M_\mathrm{UV}$ and effective radius in the UV \citep[e.g.,][]{Shibuya2019}, or more generally the so-called size-luminosity or size-mass relation. 
A larger scale-length in massive LAEs also results from suppression of UV/Ly$\alpha$ light due to more abundant dust especially in the central region \citep{Laursen2009} than in less massive LAEs. Ly$\alpha$ photons are then further affected by resonant scattering and differential extinction caused thereby, leading to $r_\mathrm{1,Ly\alpha}>r_\mathrm{1,UV}$.

On the other hand, $r_\mathrm{2,Ly\alpha}$ do not show a simple trend with respect to any photometric properties. This is again almost consistent with \citet{Leclercq2017}. This fact can be attributed to both observational and astrophysical reasons. 
First, we may simply lack the sensitivity to reveal a real trend. Even with our deep images, the Ly$\alpha$ SB profiles in Figure \ref{fig:3Lyawfit} at $r>50$ pkpc are not well constrained. 
In addition, the astrophysics involved in diffuse emission in LAHs is notoriously complicated as we discuss in Section \ref{sec:5origin}. Dominant mechanisms for Ly$\alpha$ production might be different over different mass/luminosity ranges, making a simple trend (if any) difficult to discern. 
For example, as compiled in \citet[][their Figure 7]{Kusakabe2019}, the total Ly$\alpha$ luminosity of the halo component may depend in a different way on halo mass with respect to production mechanisms (e.g., collisional excitation in cold streams vs. scattering).
Lastly, both observations \citep{Wisotzki2016, Leclercq2017} and numerical studies \citep{Lake2015,Byrohl2021} have shown that the Ly$\alpha$ SB profiles of individual LAEs are very diverse, even within galaxies with similar integrated properties. 
To conclude whether there is any trend, even larger samples and/or deeper observations are needed. With better data, we could more easily select whether two-component exponential functions or the power-law functions are preferred.

\subsubsection{Curious Behavior of Distance Subsamples: QSO Radiative History Imprinted?}\label{sec:5distance}
Subsamples based on the projected distance from the HLQSO ($d_\mathrm{Q}$) show a significant variation with a minimum at $d_\mathrm{Q}\sim10$ pMpc 
(Figure \ref{fig:3scalelp}, the second panel from right in the second row). This could just be due to the stochasticity discussed above, but if real, it could be related to the QSO's radiative history. 
$r_\mathrm{2,Ly\alpha}$ of the two largest ($r_\mathrm{2,Ly\alpha}>13$ pkpc, $d_\mathrm{Q}<6.2$ pMpc and 14.8 $<d_\mathrm{Q}<16.9$ pMpc subsample) and the smallest ($r_\mathrm{2,Ly\alpha}<7$ pkpc, $9.5<d_\mathrm{Q}<12$ pMpc subsample) differ significantly; when we repeatedly select 700 LAEs at random from the whole sample, stacking their Ly$\alpha$ images, and measuring $r_\mathrm{2,Ly\alpha}$ 1000 times, we get $r_\mathrm{2,Ly\alpha}<7$ pkpc 0.8\% of the time and $r_\mathrm{2,Ly\alpha}>13$ pkpc 12\% of the time (with the median value of $r_\mathrm{2,Ly\alpha}=10.5$ pkpc)
\footnote{If we exclude the 55 protocluster LAEs from the $d_\mathrm{Q}<6.2$ pMpc subsample, $r_\mathrm{2,Ly\alpha}$ becomes 11.3 pkpc (the $\sim30$ percentile).}.
If the HLQSO was active $\sim50$ Myrs ago, followed by $\sim30$ Myrs of inactivity, and was re-ignited $\lesssim20$ Myrs ago, the ionizing photons emitted by the QSO would have traveled a distance of $>15$ pMpc and $<6$ pMpc from the QSO
\footnote{These time estimates are lower limits calculated with projected distance and the speed of light. Propagation of ionization fronts could be delayed in some situations \citep{Shapiro2006}.}. 
These photons can ionize the envelopes of the LAEs and boost their Ly$\alpha$ luminosity, explaining the observed behavior
\footnote{see also \citet{Trainor2013,Borisova2016fluo} where the authors used QSOs associated with spectroscopic high-EW($>240$ \AA) LAEs to place limits on QSO lifetime. We see consistent behavior also in the fraction of high EW LAEs as a function of distance from the HLQSO \citep{Kida2019}. Those EW are derived with 1{\mbox{$.\!\!\arcsec$}}5 aperture. 
However, boosting EW of such a central part of LAEs with $L_\mathrm{Ly\alpha}>10^{41} \mathrm{erg~s^{-1}}$ at 16 pMpc distance would be energetically not feasible with the current HLQSO luminosity.}. 
Assuming the HLQSO 50 Myrs ago had the same luminosity as it has today (
luminosity near a rest-frame wavelength 1450 \AA, $\nu L_\mathrm{\nu,1450}=5.7\times10^{47}\mathrm{~erg~s^{-1}}$, \citealt{Trainor2012}) and isotropic radiation with escape fraction of unity, 
the ionizing radiation at 16 pMpc from the QSO can still dominate over the cosmic average UV background at $z\sim3$, $\Gamma_\mathrm{bkg}^\mathrm{z=3}=1.0\times10^{-12}\mathrm{~s^{-1}}$ \citep{Becker2013} by a factor of a few. 
\citet{Cantalupo2005} calculated the fluorescent Ly$\alpha$ emission due to QSOs in addition to the cosmic background and gave a fitting formula for an effective boost factor $b_\mathrm{eff}$ (their Equation 14--16) which can be used to estimate SB of illuminated gas clouds.
In our case, the resulting SB would be $\mathrm{SB} =(0.74+0.50(11.5(r/16\mathrm{~pMpc})^{-2}(1.6^\alpha/\alpha))^{0.89})\mathrm{SB_{HM}}$, where $r$ is the distance from the HLQSO in pMpc, $\alpha$ is the QSO's spectral slope ($L_\nu\propto\nu^{-\alpha}$), and $\mathrm{SB_{HM}}=3.67\times10^{-20}\mathrm{~erg~s^{-1}~cm^{-2}~arcsec^{-2}}$ is the expected SB without QSO boost. 
Assuming $\alpha\sim1$ gives $\mathrm{SB}=2.7\times10^{-19}\mathrm{~erg~s^{-1}~cm^{-2}~arcsec^{-2}}$.
Thus, QSO-induced fluorescence is energetically possible to have caused variation in $r_\mathrm{2,Ly\alpha}$ in the projected distance subsamples, at least in the optimistic case.
In reality, our narrow-band selection picks up LAEs with line-of-sight distance uncertainty of $\sim19$ pMpc, which is the same level as the radius of the FoV of our images, and this randomizes light-travel time from the QSO to each LAE. 
Such effects further complicate the situation, but we have shown that under some circumstances with appropriate QSO light curve and line of sight distribution of LAEs, the observed trend of $r_\mathrm{2,Ly\alpha}$ might be explained.
Upcoming instruments such as Prime Focus Spectrograph \citep[PFS; a wide-field multi-fiber spectrograph, ][]{Takada2014} on the Subaru Telescope can test this fluorescence scenario by obtaining systemic redshifts of LAEs and thus reducing uncertainties on their 3D distances from the QSO.

\subsubsection{Origin of the Large LAH of Protocluster LAEs}\label{sec:5environ}
We showed that the dependence of LAHs on environment is not large where $\delta<2.5$, but in the protocluster ($\delta>2.5$) the LAHs show elevated flux out to $>100$ pkpc (Figure \ref{fig:3scalelp}, \ref{fig:Lyaprofs2}). 
If LAHs trace the gas distribution of the CGM, the former indicates that the large-scale environment (except for protocluster environments) does not have a large impact on the matter distribution out to $\sim100$ pkpc and it is determined by individual halo mass or other internal processes. 
Alternatively, LAEs could simply be poor tracers of large-scale environments. Recently, \citet{Momose2021} found that LAEs behind the foreground large-scale structure tend to be missed due to absorption by the foreground structure \citep[see also][]{Shimakawa2017}. 
It may also be the case that even our large field of view of 1.2 deg diameter may be insufficient to capture diverse environments including voids while targeting a single protocluster. 
Other line emitters or continuum-selected galaxies would be ideal, though would be expensive for current facilities. 
On the other hand, the HS1549 protocluster is confirmed by overdensity of LAEs and continuum-selected galaxies, with $\sim200$ member galaxies spectroscopically identified (\citealt{Trainor2012,Mostardi2013}; C. Steidel et al. in prep.). 
Enlarging the size of cutout images, we confirmed that flux higher than $>1\sigma$ level continues out to $\sim500$ pkpc. In the following, we investigate the possible cause of this emission in the $\delta>2.5$ subsample.

\begin{figure}
\includegraphics[angle=270,width=\columnwidth]{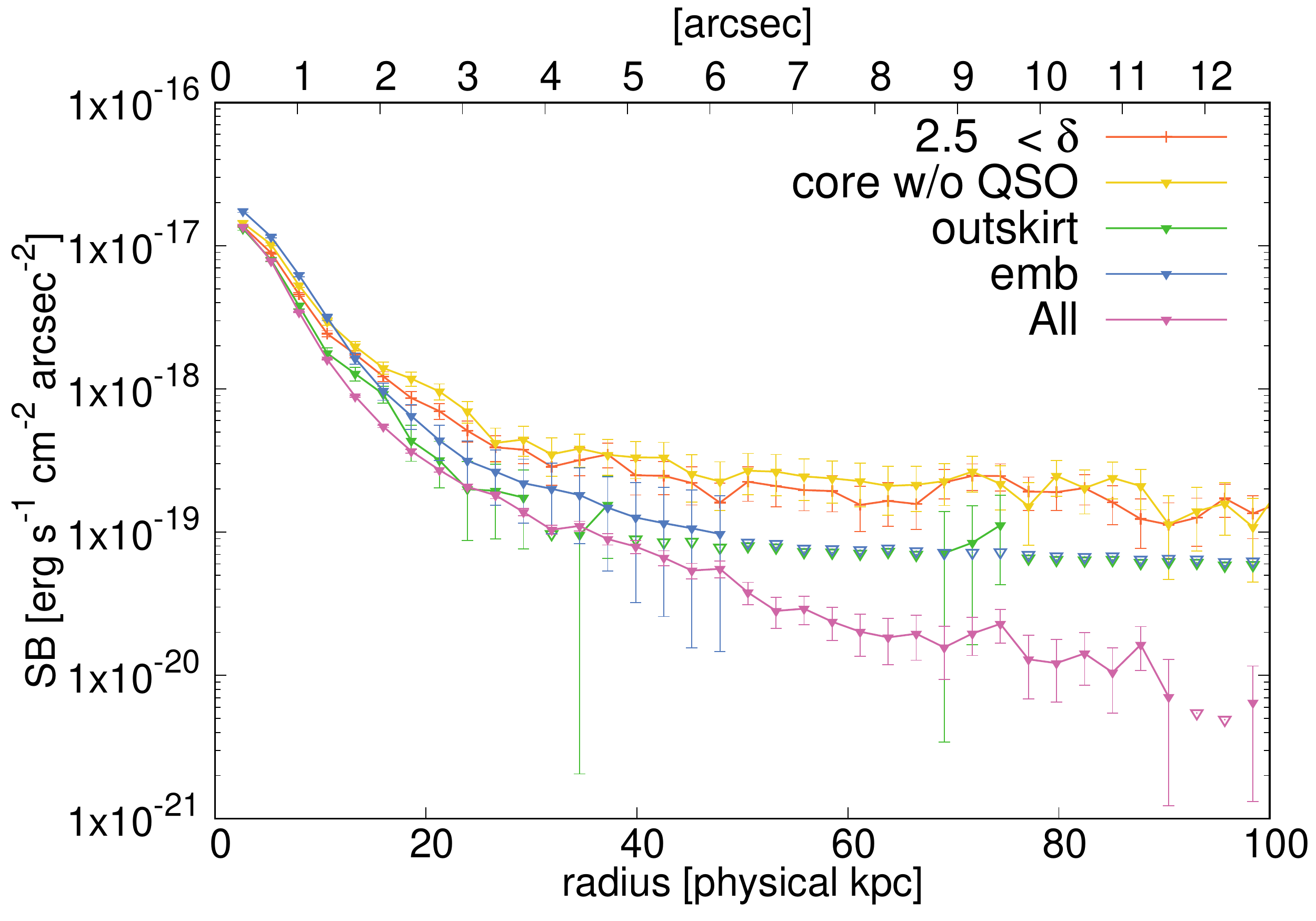}
  \caption{Stacked Ly$\alpha$ SB profiles of the protocluster ($\delta>2.5$) sample (orange), the core sample (yellow), the outskirt sample (green), embedded and stacked image (blue, see text), and all LAEs (purple, same as Figure \ref{fig:3Lyaall}). Downward triangles show 1$\sigma$ error levels after residual sky subtraction.
\label{fig:3nearfar}}
\end{figure}

First, we further divide the protocluster sample into a ``core'' group (LAEs within a projected distance of $<500$ pkpc from the HLQSO but excluding the HLQSO itself, $N=25$) and an ``outskirt'' group (the remainder, $N=29$), and stacked them separately. 
This time, pixels with SB$> 10^{-17} \mathrm{~erg~s^{-1}~cm^{-2}~arcsec^{-2}}$ around the HLQSO and the associated bright nebula \citep{Kikuta2019} are masked before stacking in each cutout image to exclude their contribution. As shown in Figure \ref{fig:3nearfar}, the core sample (the yellow curve) clearly shows excess emission which does not decrease toward the edge above the original orange curve, while the outskirt stack (the green curve) shows just a mild bump around $\sim15$ pkpc and no evidence of excess. 
This suggests the extended emission of $\delta>2.5$ sample is solely produced by the LAEs at the core of the protocluster. 

Recalling that the protocluster LAEs tend to have higher $M_\mathrm{UV},~L_\mathrm{Ly\alpha}$ and lower EW (Figure \ref{fig:cumdis0}) and that such galaxies generally have more extended Ly$\alpha$ SB profiles (Figure \ref{fig:3scalelp} and Figure \ref{fig:Lyaprofs1}), the overabundance of such LAEs leads to larger LAHs. 
Moreover, the core region is sufficiently crowded that Ly$\alpha$ emission from neighboring LAEs may overlap, thereby leading to an overestimation of the SB profile. 
We evaluated this effect using the stacked Ly$\alpha$ images of subsamples based on UV magnitude (those shown in the top row of Figure \ref{fig:3Lyaimage1}) as follows. We embedded their images in a blank image, mimicking the observed spatial distribution of LAEs (including the HLQSO) using IRAF task ``mkobjects''. 
When embedding LAEs with a certain $\mathrm{M_{UV}}$, we used the stacked Ly$\alpha$ image of the appropriate $\mathrm{M_{UV}}$ subsample scaled to match the observed UV magnitude. 
For example, we embedded the scaled stacked image of the $-19.2<\mathrm{M_{UV}<-18.6}$ subsample at the location of LAEs in the same UV magnitude range. 
Cutout images of the simulated Ly$\alpha$ image were then created at the locations of the embedded LAEs and stacked in the same manner as the real ``core'' LAEs. 
The result is also shown in Figure \ref{fig:3nearfar} (the blue curve). The observed large LAH (the orange curve) is not reproduced. 
Thus, we conclude that the combined effect of the overabundance of bright LAEs and overlapping do contribute to the large LAHs, but cannot fully explain the extent of the LAH of the core LAEs.

\citet{Kikuta2019} showed that there is a Mpc-scale diffuse Ly$\alpha$ emitting structure around the HLQSO. Such diffuse emission would explain the remaining excess in the core region of the protocluster. 
The excess comes not only from gas directly associated with LAEs but also from gas out of the LAEs; the pixel value distribution of the core region ($<500$ pkpc from the HLQSO) after masking $<10$ arcsec ($\sim80$ pkpc) regions around the detected LAEs and a $\sim200$ pkpc $\times 260$ pkpc box covering the bright QSO nebula clearly shows excess at $1\times10^{-18}<$SB$<4\times10^{-18} \mathrm{~erg~s^{-1}~cm^{-2}~arcsec^{-2}}$ compared to that of the outer region (with regions around LAEs masked in the same manner as the core case). 
Previously, two studies reported similar large diffuse LAHs in protoclusters at $z=2$--3; \citet{Steidel2011} observed HS1549 ($z=2.84$, same as this study
\footnote{Although their LBG sample is an order of magnitude brighter in UV than our LAE sample, we did a consistency check with \citet{Steidel2011} results; we confirmed that our stack of LBGs used also in their sample gives a consistent Ly$\alpha$ SB profile as their work.}), 
HS1700 \citep[$z=2.30$; see also][]{Erb2011}, and SSA22 \citep[$z=3.09$; also observed by][]{Matsuda2012}.
For the HS1549 and SSA22, direct evidence of Mpc-scale diffuse Ly$\alpha$ emission has been observed (\citealt{Kikuta2019} and \citealt{Umehata2019}, respectively), and the identification of filamentary structure traced by six Ly$\alpha$ blobs in the HS1700 protocluster by \citet{Erb2011} suggests that this protocluster also harbors such diffuse emission.
In the forming protocluster core at $z=2$--3, a large amount of cold gas can be accreted through the filamentary structure (the cosmic web) penetrating the cores \citep{Keres2005,Keres2009,Dekel2009}. 
In addition to abundant gas, \citet{Umehata2019} showed that an enhanced ionizing UV background due to a local overdensity of star-forming galaxies and AGNs may play a crucial role in boosting fluorescent Ly$\alpha$ emission to a detectable level.
The HS1549 protocluster also has tens of active sources (i.e., AGN, Ly$\alpha$ blobs \citep[][C. Steidel et al. in prep.]{Kikuta2019}, and SMGs \citep{Lacaille2019}) within a few arcmins from the HLQSO, producing sufficient UV radiation to power the diffuse emission; in Supplementary Material S9 of \citet{Umehata2019}, they calculated the required UVB strength to boost SB of optically thick gas. Fifty times stronger UVB than the cosmic average at $z=3$, which is easily realized by the HLQSO alone in the area within several pMpc from it, would boost SB to $\sim10^{-18} \mathrm{~erg~s^{-1}~cm^{-2}~arcsec^{-2}}$ level. 
To summarize, the very large LAHs of stacked Ly$\alpha$ profiles reported previously and in this work can be attributed to an overlap of crowded LAEs and diffuse fluorescent Ly$\alpha$ emission within the forming protocluster core. The $\delta^2$ dependence of LAH scale-length claimed in \citet{Matsuda2012} should be revisited using new data targeting more protocluster fields at similar redshift together with appropriate analysis methods as discussed in Section \ref{sec:5prev}.

\subsection{Discovery of ``UV Halos'' and Its Implication to Low-Mass Galaxy Evolution}\label{sec:5uv}

As seen in Figure \ref{fig:3uvwfit} and Figure \ref{fig:3scalelp} in Section \ref{sec:ressub}, we have discovered ``UV halos'' around UV/Ly$\alpha$-bright and/or low-EW LAEs. This has a significant impact on our understanding of the origin of LAHs, and also of galaxy evolution, because it provides direct evidence of star formation activity in the outskirts of high-redshift low-mass galaxies. 

\begin{figure}
\plotone{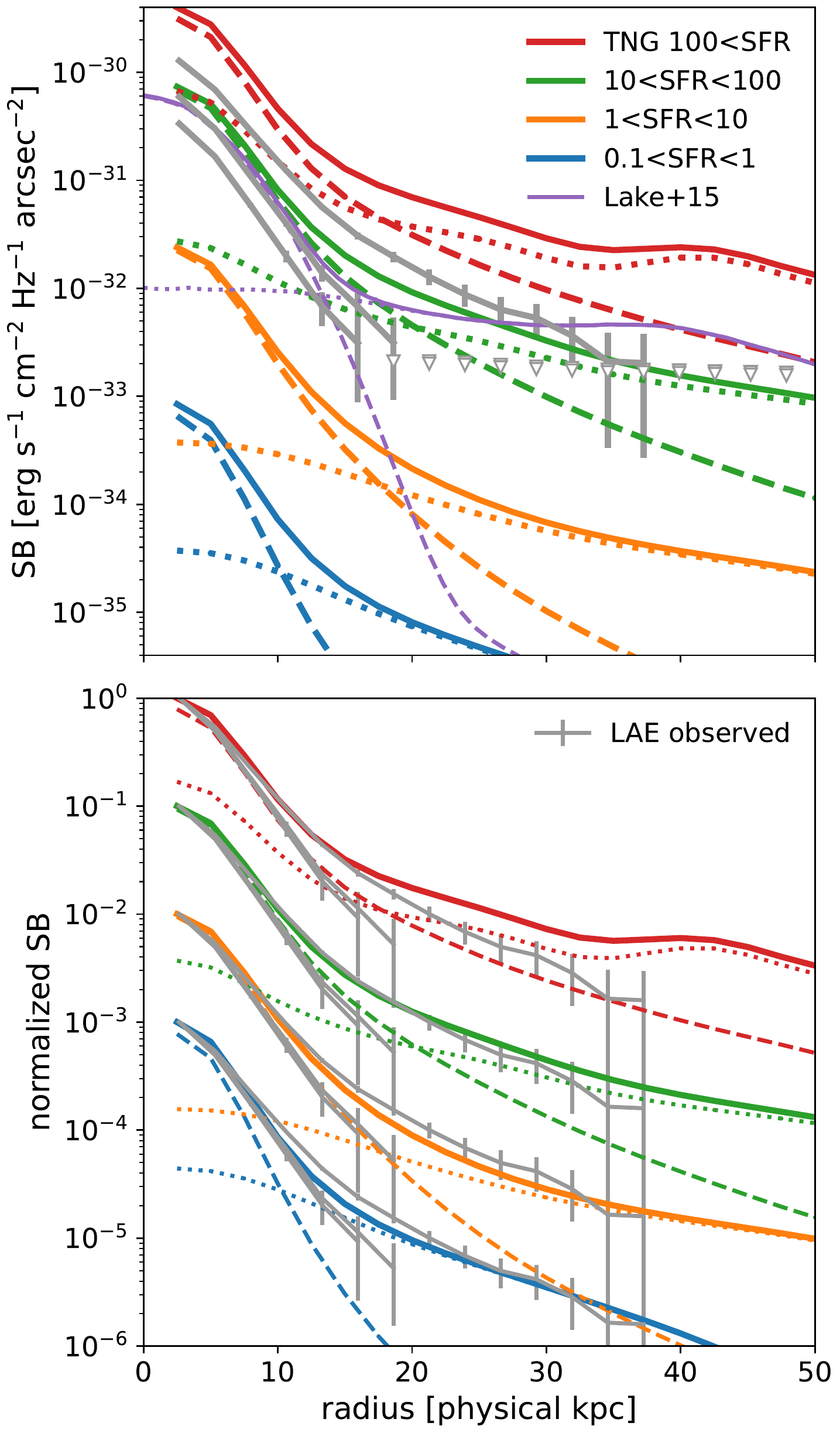}
  \caption{
  (Top) SFR surface density profiles of TNG100 galaxies converted to UV SB profiles (thick colored solid curves, see text). Different colors show different SFR bins. Dashed and dotted curves show contributions to the profiles from the main halo and other halos, respectively. The thin purple curve shows that predicted in \citet{Lake2015}. Gray curves with error bars are observed UV SB profiles of our three UV-brightest LAE subsamples. These profiles are corrected for a slight redshift dimming effect by scaling $(1+z)^3/(1+2.845)^3$.
  (Bottom) Curves of top panels are normalized to better compare their shapes.
\label{fig:5tnguv}}
\end{figure}

To gain more insight on the latter point, we used the data products from the TNG100 run of the IllustrisTNG simulation \citep[e.g.,][]{Nelson2018,Pillepich2018}. 
We make median stacked SFR surface density profiles of FOF (friends-of-friends) halos, which roughly represent collections of gravitationally bound DM particles, at $z=3$ for 4 SFR bins; $0.1<\mathrm{SFR}<1~\mathrm{M_\odot~yr^{-1}}$, $1<\mathrm{SFR}<10~\mathrm{M_\odot~yr^{-1}}$, $10<\mathrm{SFR}<100~\mathrm{M_\odot~yr^{-1}}$, and $100~\mathrm{M_\odot~yr^{-1}}<\mathrm{SFR}$ and compare them with those of our three UV-brightest subsamples after converting the simulation data to UV flux density using the SFR-UV luminosity density conversion of \citet{Murphy2011,Kennicutt2012} relation (Figure \ref{fig:5tnguv}). 
The SFR surface density profiles were convolved with a Gaussian with 0{\mbox{$.\!\!\arcsec$}}77 FWHM\footnote{As we see in Section \ref{sec:analyses}, the PSF of our images are not Gaussian but have a power-law tail with an index of $\sim-3$, but the PSF behavior at large radii does not affect our result.} 
to make a fair comparison. 
Here, SFR of TNG galaxies denotes a total of all particles which belong to one FOF halo. 
We also plot the prediction of \citet[][convolved with a Gaussian with 1{\mbox{$.\!\!\arcsec$}}32 FWHM or 10.3 pkpc at $z=3.1$]{Lake2015} for discussion in Section \ref{sec:5origin}.
The three UV-bright subsamples have median UV absolute magnitude of $-19.62, -18.88, -18.31$ (Table \ref{tab:halostack}), but these were derived using 1{\mbox{$.\!\!\arcsec$}}5 diameter apertures and thus may be underestimated. The total magnitudes derived by integrating the UV SB profiles down to the radii where emission is detected at more than $1\sigma$ significance are $-21.34, -20.49, -20.10$, respectively, corresponding to SFR of $17, 7.7, 5.4~\mathrm{M_\odot~yr^{-1}}$. 
The SFR of simulated galaxies whose UV SB profiles match those of our LAEs is higher than that of our LAEs (10--100 vs. 5.4--17), 
but there is considerable uncertainty in the conversion between UV luminosity and SFR, as it depends on stellar age, dust attenuation, and metal abundance, and not all galaxies in simulations would be selected as LAEs.
Reconciling this mismatch is beyond the scope of our paper.
Rather, to compare the profile shapes we normalized each curve in the bottom panel of Figure \ref{fig:5tnguv}. 
While the UV SB profiles of the two fainter subsamples seem to be slightly more compact than the SFR density profiles of TNG galaxies,
the UV-brightest subsample has a remarkably similar shape as the SFR surface density profiles of TNG galaxies with $1<\mathrm{SFR}<10$ and $10<\mathrm{SFR}<100$ subsamples.

\begin{figure*}
\includegraphics[width=\textwidth]{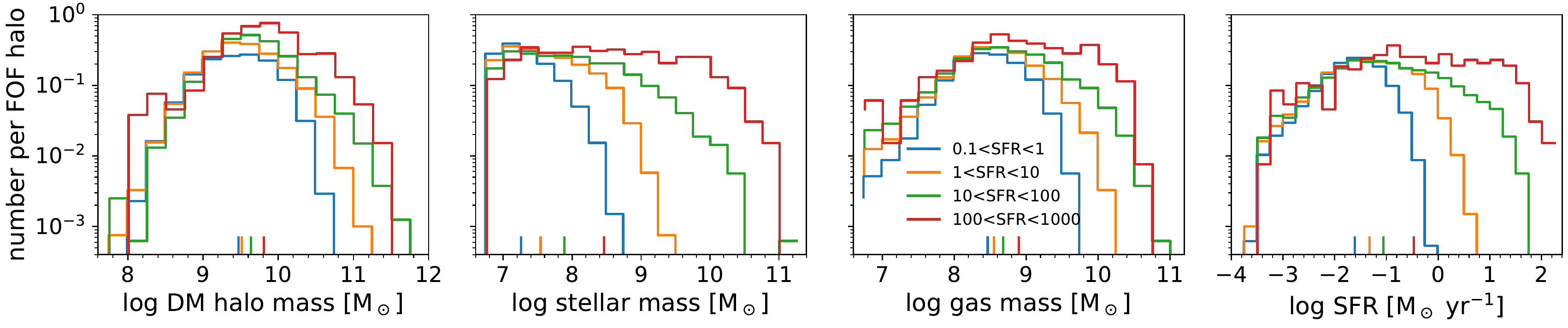}
  \caption{
  The distribution of DM mass, stellar mass, gas mass, and SFR (from left to right) of satellite halos for different main halo's SFR bins, the corresponding colors to which are shown in the third panel. Small colored ticks below each panel indicate the medians of each bin. 
\label{fig:5tnghalos}}
\end{figure*}

We further decompose the simulated FOF halos into the main halo and subhalos. 
The decomposed SFR profiles demonstrate that the flattened outer part is dominated by the contribution of satellites. 
Given the similarity of the profiles, we suggest that the UV halo of the UV-brightest LAEs is also due to such satellites. 
To characterize the subhalos around the central galaxy, we extracted dark matter halo mass, gas mass, stellar mass, and SFR of those within 50 pkpc ($\sim6$ arcsec, 2D projected distance) from their main halos. 
We only handle those halos with $M_\mathrm{DM}>7.5\times10^7~\mathrm{M_\odot}$ ($=10\times$DM particle mass), $M_\mathrm{stellar}>7.0\times10^6~\mathrm{M_\odot}$ ($=5\times$stellar particle mass), and non-zero SFR to avoid spurious objects. 
The distribution of DM mass, stellar mass, gas mass, and SFR of satellites are plotted in Figure \ref{fig:5tnghalos}.
Satellites responsible for the UV halo of our LAEs would be similar to those around central galaxies with $1 < \mathrm{SFR} < 10$ and $10 < \mathrm{SFR} < 100$. 
On average, they have 1.9 and 2.3 satellites, respectively, with median DM halo masses of $3.3\times10^9~\mathrm{M_\odot}$ and $4.4\times10^9~\mathrm{M_\odot}$, and mean total halo SFR of 0.30 and 2.6 $\mathrm{M_\odot}~\mathrm{yr^{-1}}$ (i.e., $\sim10\%$ of central galaxies). 
This suggests that the UV halo is comprised of a few satellite galaxies around the 
main halo and not by an intrinsically diffuse halo, unlike optical stellar halos of local galaxies \citep[e.g.,][]{DSouza2014}.
Under this hypothesis, individual LAEs would not have smooth UV SB profiles like those presented in Figure \ref{fig:3uvwfit}, but would be more likely to exhibit stochastic shapes made by $\sim2$ discrete satellites; thus the term UV ``halo'' may not be appropriate, if the quoted simulations represent the reality. 
The smooth profile of stacked UV SB is simply a reflection of the radial and SFR distribution of satellite galaxies.

Are such satellites observable? 
The SFR distributions of satellites around galaxies with $0.1<\mathrm{SFR}<10$ in the simulations have medians in the range $0.01<\mathrm{SFR / [M_\odot~\mathrm{yr^{-1}}]}<0.1$, although some very low-mass objects may be affected by resolution effects (Figure \ref{fig:5tnghalos}). 
This translates into UV absolute magnitude of $-13.3$ and $-15.8$, and apparent g-band magnitude of 32.1 to 29.6 mag at $z=2.84$ (assuming no K-correction, or equivalently, flat UV continua).
Brighter satellites are thus well within the reach of HST and JWST sensitivity in some deep fields or with the aid of gravitational lensing \citep[e.g.,][]{Alavi2016,Bouwens2022a}.

\subsection{On the Origin of the LAHs \label{sec:5origin}}
Lastly, we infer the origin of LAHs of LAEs through comparison with recent numerical simulations. 
As introduced in Section \ref{sec:intro}, Ly$\alpha$ photons in the halo regions are generated either by ex-situ (mostly from the host galaxy) and transported by resonant scattering in neutral gas, or in-situ via photoionization followed by recombination or collisional excitation. 
In-situ photoionization is maintained by ionizing photons from star formation and/or AGN activity in satellites, central galaxies, or other nearby sources of ionizing UV such as QSOs. In-situ collisional excitation would be driven by shocks due to fast outflows due to feedback or gravitational energy of inflowing gas, but the former is considered to be effective in more massive and energetic sources such as radio galaxies and Ly$\alpha$ blobs \citep[e.g.,][]{Mori2004}. 
In-situ Ly$\alpha$ photons may also experience resonant scattering, but its effect on the redistribution of photons is likely to be relatively weak due to lower HI column densities than the central part by more than a few dex \citep{Hummels2019,VandeVoort2019}.

To predict Ly$\alpha$ SB profiles around galaxies, one needs to know physical parameters such as hydrogen density, neutral fraction, gas kinematics, and temperature which depend on SFR and AGN activity around a point of interest, and ionizing/Ly$\alpha$ photon escape fraction, etc. 
Solving all these quantities is practically impossible, but recent simulations are beginning to reproduce observed Ly$\alpha$ SB profiles reasonably well. 
Among such studies is \citet[][hereafter B21]{Byrohl2021}; B21 presents full radiative transfer calculations via post-processing of thousands of galaxies in the stellar mass range $8.0<\log(M_*/M_\odot)<10.5$ drawn from the TNG50 simulations of the IllustrisTNG project. 
One of the advantages of B21 is the sample size, which is far larger than those of previous studies.
For example, \citet[][hereafter L15]{Lake2015} performed radiative transfer modeling of 9 LAEs, obtaining significantly diverse SB profiles. 
Those with massive neighbors have elevated SB profiles both in UV and Ly$\alpha$, significantly boosting the average profile. 
But the 9 galaxies modeled in the simulation would not be representative of star-forming galaxies at $z\sim3$ if not carefully selected. 
\citet{Mitchell2020a} calculated Ly$\alpha$ SB profiles of a single galaxy at $z=3$--4 using the RAMSES-RT code, a radiation hydrodynamics extension of the RAMSES code, and succeeded in reproducing SB profiles similar to MUSE observations. The small sample size is somewhat mitigated by using all available outputs between $z=4$ and $z=3$, but still there may remain biases with respect to environment or evolutionary phase. 
For these reasons, we compare our results primarily with those of B21.

\subsubsection{Dominance of Star Formation in Central and Satellite Galaxies}\label{sec:5centsat}
\begin{figure}[h]
  \begin{center}
\includegraphics[width=\columnwidth]{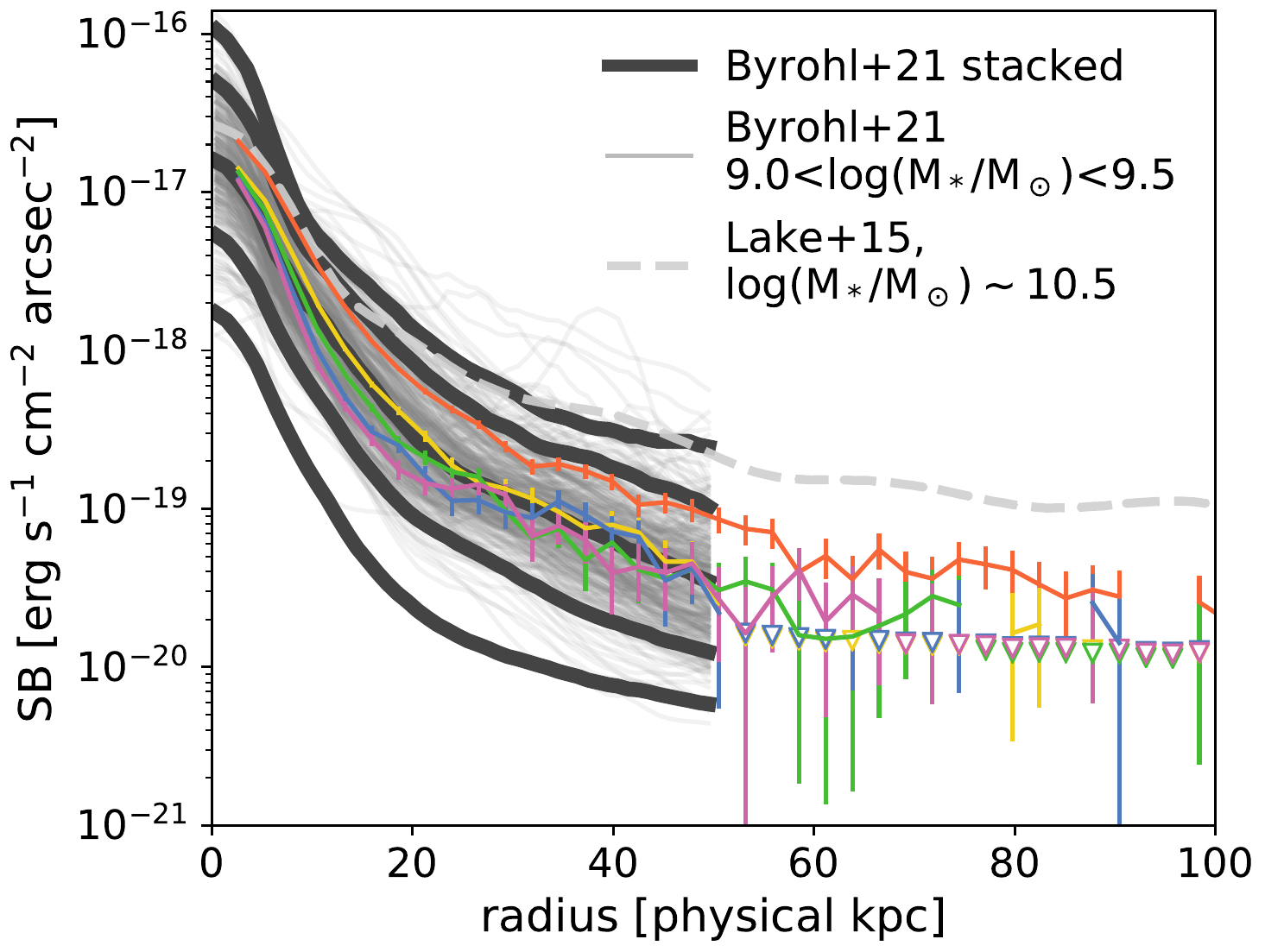}
  \caption{
  Stacked Ly$\alpha$ SB profiles of galaxies at $z=3$ for five different intervals of stellar mass from the \citet{Byrohl2021} simulation (thick gray) along with the averaged SB profile predicted in \citet{Lake2015} (dashed) and our results (colored with errorbars, UV subsamples from the top left panel of Figure \ref{fig:Lyaprofs1}). The thick gray curves represent, from bottom to top, profiles of galaxies with $8.0<\log(M_*/\mathrm{M_\odot})<8.5$, $8.5<\log(M_*/\mathrm{M_\odot})<9.0$, $9.0<\log(M_*/\mathrm{M_\odot})<9.5$, $9.5<\log(M_*/\mathrm{M_\odot})<10$, and $10<\log(M_*/\mathrm{M_\odot})<10.5$. 
  To show the diversity of Ly$\alpha$ SB profiles, we show all profiles of galaxies with $9.0<\log(M_*/\mathrm{M_\odot})<9.5$ with thin gray curves. 
  The B21 and L15 results are corrected for a slight redshift dimming effect by scaling $(1+z)^4/(1+2.845)^4$ ($z=3.0$ for B12 and 3.1 for L15).
\label{fig:5byrohl}}
  \end{center}
\end{figure}

In Figure \ref{fig:5byrohl}, we plot B21 (convolved with a Gaussian with 0{\mbox{$.\!\!\arcsec$}}7 FWHM) and L15 (convolved with a Gaussian with 1{\mbox{$.\!\!\arcsec$}}32 FWHM) results with our observations of subsamples based on UV magnitude. 
Although B21 probed only up to $<50$ pkpc, the overall shapes of the $9.5<\log(M_*/M_\odot)<10.0$ stack and our UV-brightest subsample stack match remarkably well, and the $9.0<\log(M_*/M_\odot)<9.5$ stack and the other UV subsamples stack also show fairly good agreement. 
We also highlight the diversity of Ly$\alpha$ SB profiles here by drawing all the profiles of galaxies with $9.0<\log(M_*/M_\odot)<9.5$ in B21 in Figure \ref{fig:5byrohl} with thin curves. 
Similarly to L15, bumps in some curves are caused by companion galaxies. 
This demonstrates the difficulty of studying halo origins with a small sample as discussed in Section \ref{sec:5dependence}. 
Future (observational and theoretical) studies should keep this in mind before discussing the halo dependence on physical properties.
B21 concluded that scattering of Ly$\alpha$ photons originate from star formation in the central and satellite galaxies is almost always dominant within 50 pkpc from the center ($\sim50$\% at $> 20$ pkpc), with in-situ collisions and recombination explaining the remaining 30\% and 20\%, respectively. 
The contribution from satellite star formation dominates over that from the central galaxy beyond $\sim40$ pkpc, and they show that halos that have more massive neighbors within 500 pkpc can have very extended LAHs compared to those residing in normal environments (Figure 12 of B21). 
Similar conclusions are reached by L15 and \citet{Mitchell2020a} as to the importance of satellites. 
The dashed curve shows the L15 result which extends to 100 pkpc. As we saw in Figure \ref{fig:5tnguv} (see also Figure 4 of L15), their galaxies (whose mean stellar mass is $\sim2.9\times10^{10}\mathrm{~M_\odot}$) have more star formation outside of the host halos and have an enhanced SB profile at outer regions. A comparably massive galaxy sample is required to confirm their prediction. 
Our first detection of satellites (Section \ref{sec:5uv}) and reasonable agreement of both UV (Figure \ref{fig:5tnguv}) and Ly$\alpha$ (Figure \ref{fig:5byrohl}) SB profiles with simulations suggest that star formation in central and satellite galaxies are important Ly$\alpha$ sources contributing to LAHs.

Rest-frame Ly$\alpha$ EW of each annulus of observed LAEs is plotted with the right axis of Figure \ref{fig:3uvwfit}. 
Considering the outward diffusion of Ly$\alpha$ from the central galaxies, this EW is always an upper limit for the EW of in-situ Ly$\alpha$ emission. 
If low-mass satellites are responsible for the outer LAHs as simulations suggest, then their expected dark matter halo masses are about $10^{9-10}\mathrm{~M_\odot}$. 
EW$_\mathrm{0,Ly\alpha}$ of $\gtrsim200$ \AA~ observed at $r\sim30$ pkpc of the UV/Ly$\alpha$-brightest or EW-lowest subsamples can be explained by halo star formation alone if these low-mass galaxies have average EW$_\mathrm{0,Ly\alpha}$ of $\geq200$ \AA, even without scattered Ly$\alpha$ from central galaxies. 
For other subsamples, the EW$_\mathrm{0,Ly\alpha}$ is $>240$ \AA~ due to the faintness of UV SB and the extended Ly$\alpha$ SB profiles at outer regions. Such high EW$_\mathrm{0,Ly\alpha}$ is hard to explain by star formation alone \citep{Schaerer2003a}; scattering of Ly$\alpha$ photons produced elsewhere and in-situ recombination/collisional excitation should dominate the Ly$\alpha$ photon budget.

\subsubsection{Non-Negligible Contribution from In-Situ Ly$\alpha$ Production}
Processes other than star formation, e.g. QSO-boosted fluorescence or collisional excitation via gravitational cooling, can still be important at large radii, since they are predicted to make non-negligible contributions and there are situations under which such processes become more important. 
In Section \ref{sec:5distance} and Section \ref{sec:5environ}, we specified conditions where fluorescence could be dominant, namely regions near bright QSOs and/or near protocluster cores. 
A number of simulations have suggested that stronger ionizing radiation fields boost fluorescent Ly$\alpha$ emission from the CGM/IGM \citep[][see also Appendix A5 of B21]{Cantalupo2005,Kollmeier2010}.
A major problem is that both processes are very hard to accurately predict even with state-of-the-art simulations; recombination emissivity could be significantly boosted without changing total hydrogen column density if there are many tiny ($\ll1$ pkpc) clumps with locally increased density in the CGM/IGM regions unresolved in current standard simulations but suggested from observations \citep{Rauch1999,Cantalupo2014,Cantalupo2019,McCourt2018,Hummels2019,VandeVoort2019}. 
The total Ly$\alpha$ luminosity from gravitational cooling should have a strong dependence on halo mass \citep{Goerdt2010}.
In addition, the emissivity of the collisional process depends extremely sensitively on temperature exponentially in the range $T=10^4$--$10^5$ K characteristic of cold accretion, and treatments of the effect of self-shielding against the UVB may have a critical impact on results \citep{Faucher-Giguere2010,Kollmeier2010,Rosdahl2012}. 

There remains a possibility that our main conclusion about the dominance of central/satellite star formation may apply only to relatively massive LAEs, since lower-mass halos would have less scattering media and less satellites. 
For example, high-EW LAEs are efficient producers of Ly$\alpha$ photons. 
But because they are on average less massive and should have lower HI gas \citep{Rakic2013,Turner2017}, they have more compact LAHs despite efficient Ly$\alpha$ production. 
\citet{Kakiichi2018} showed that scattering of Ly$\alpha$ photons produced by central galaxies with realistic HI distributions constrained by Ly$\alpha$ forest observations of LBGs results in power-law-like Ly$\alpha$ SB profiles \citep[see also][]{Steidel2011}. 
In Figure \ref{fig:3Lyawfit}, UV/Ly$\alpha$-faint and high-EW LAEs seem to deviate from power-law fits at $r>25$ kpkc. This could be a hint of the dominance of other processes. 
Similar conclusion of the dominance of scattered Ly$\alpha$ from central galaxies and possible contribution from the other processes is reached by recent observational studies \citep{LujanNiemeyer2022,LujanNiemeyer2022a}. 
In this way, much information is buried beyond the ``flattening radius'' around 20 pkpc, outside of which contributions from central/satellites appear insufficient to explain the observations. 
With a larger sample and deeper data, we can further probe the behavior of LAHs e.g., by making EW-based subsamples with matched UV magnitude, etc.

\section{Summary}\label{sec:summary}
We have investigated the rest-frame UV continuum ($\lambda\sim1225$ \AA) and Ly$\alpha$ radial surface brightness (SB) profiles of LAEs at $z=2.84$ through stacking analyses of UV and Ly$\alpha$ images created from Subaru/HSC g-band and the NB468 narrow-band images. 
The depth and wide field coverage, including a known protocluster, enable us to study both SB profiles with unprecedented depth because of the large sample ($N=3490$) at $z\sim3$. Our major findings are as follows:
\begin{enumerate}
    \item Stacking of 3490 LAEs yields a SB sensitivity of $\sim1\times10^{-20}\mathrm{~erg~s^{-1}~cm^{-2}~arcsec^{-2}}$ in Ly$\alpha$ and $\sim1\times10^{-33}\mathrm{~erg~s^{-1}~cm^{-2}~Hz^{-1}~arcsec^{-2}}$ (Figure \ref{fig:3Lyauvwfit}). 
    Our analyses reveal that systematic errors should be at the same levels at most (Section \ref{sec:4systematics} and that the choice of mesh size for local sky estimation could have a large impact on the results (Figures \ref{fig:3Lyaall} and  \ref{fig:3Lyaskymesh}). 
    \item By dividing the LAEs into subsamples according to various photometric properties, UV magnitude, Ly$\alpha$ luminosity, Ly$\alpha$ EW, projected distance from a hyperluminous (HL) QSO residing at the center of the protocluster (as a proxy for radiation field strength boosted by the HLQSO), and LAE overdensity $\delta$ on a 1{\mbox{$.\!\!\arcmin$}}8 ($\sim840$ pkpc) scale, we study the dependence of the SB profiles on these quantities (Figures \ref{fig:3Lyawfit} and \ref{fig:3uvwfit}). 
    To quantify the radial dependence of SB profiles, we fit 2-component exponential functions (Equation \ref{eq:exp}) to observed profiles. 
    For Ly$\alpha$ SB profiles, we consistently obtain $r_\mathrm{2,Ly\alpha}$, the scale-length of the more extended component, of $\sim10$ pkpc for all subsamples. 
    However, we do not observe any clear trend of $r_\mathrm{2,Ly\alpha}$ with any property probed here (Figure \ref{fig:3scalelp}), whereas the scale-length of the compact components (both of $r_\mathrm{1,Ly\alpha}$ and $r_\mathrm{1,UV}$) varies monotonically with respect to UV magnitude, Ly$\alpha$ luminosity, and Ly$\alpha$ EW.
    \item We find an exceptionally large exponential scale-length $r_\mathrm{2,Ly\alpha}$ for LAEs in the inner core (those within 500 pkpc from the HLQSO) of the protocluster, and a significant variation in $r_\mathrm{2,Ly\alpha}$ with respect to the projected distance from the HLQSO. 
    These findings could be explained by enhanced ionizing background radiation due to abundant active sources and cool gas at the forming protocluster core and the past activity of HLQSO, respectively (Section \ref{sec:5distance} and Section \ref{sec:5environ}).
    \item We for the first time identify extended UV components (i.e., $r_\mathrm{2,UV}$ inconsistent with zero), or ``UV halos'' around some bright LAE subsamples, which provides direct evidence for the contribution of star formation in halo regions and/or satellite galaxies to Ly$\alpha$ halos. Comparison with cosmological hydrodynamical simulations suggests that UV halos could be composed of 1--2 low-mass ($M_\mathrm{DM}\sim10^{9.5}~\mathrm{M_\odot}$) galaxies (Section \ref{sec:5uv}) with total SFR of $\sim10\%$ of that of their central galaxies. 
    \item Combining our results with predictions of recent numerical simulations, we conclude that star formation in both the central galaxy and in satellites, together with resonant scattering of Ly$\alpha$ photons are the dominant factors determining the Ly$\alpha$ SB profiles at least within a few tens pkpc. In outer regions (projected distances $\gtrsim30$ pkpc) other mechanisms such as fluorescence can also play a role especially in certain situations like dense regions of the Universe and near zones of bright QSOs (Section \ref{sec:5origin}).
\end{enumerate}

The low-mass satellite galaxies suggested by our deep UV stacked images will be very important targets for revealing the role of minor mergers in galaxy evolution and cosmic reionization, as they are believed to be promising analogues of the main galaxy contributors to ionizing photon budget at $z>6$ \citep{Robertson2013,Mason2015}. 
In a year or so, deep surveys with JWST will detect these galaxies within several arcsecs from LAE-class galaxies, which are to be observed in coming programs. 
Finally, H$\alpha$ observations of LAHs open up a new pathway to study star formation in the CGM and fluorescent clumps without the blurring effect of resonant scattering.
Observations of H$\alpha$ from $z=2.84$ has not been possible with ground-based telescopes due to heavy atmospheric absorption and extremely bright backgrounds, but now this is also becoming feasible thanks to JWST. 
In the future, we will combine cross-analyses with a larger LAE sample and new constraints from JWST to further constrain the origin of LAHs. 

\acknowledgments
We are grateful to the anonymous referee for their careful reading and constructive comments and suggestions, Chris Byrohl for providing simulation data, and Haibin Zhang, Masafumi Yagi, Masayuki Umemura, Tadafumi Takata, Kazuhiro Shimasaku, Yusei Koyama, and Kazuhiro Hada for fruitful discussions. 
We thank Yukie Oishi and the HSC pipeline team for their helpful comments on HSC data analyses. 
We would like to acknowledge all who supported our observations at the Subaru Telescope, including the staff of the National Astronomical Observatory of Japan, Maunakea Observatories, and the local Hawaiian people who have been making efforts to preserve and share the beautiful dark sky of Maunakea with us. 
We are honored and grateful for the opportunity of observing the Universe from Maunakea, which has the cultural, historical, and natural significance in Hawaii. 
Data analysis was carried out on the Multi-wavelength Data
Analysis System operated by the Astronomy Data Center (ADC), National Astronomical Observatory of Japan and on analysis servers at Center for Computational Astrophysics, National Astronomical Observatory of Japan. 
SK acknowledges supports from the Japan Society for the Promotion of Science (JSPS) KAKENHI grant Nos. 18J11477, 19H00697 and the Course-by-Course Education Program of SOKENDAI. YM acknowledges supports from the JSPS KAKENHI grant Nos. 25287043, 17H04831, 17KK0098. 
CCS acknowledges support by US NSF grant AST-2009278.
ZZ acknowledges support by US NSF grant AST-2007499.

\vspace{5mm}
\facilities{Subaru (HSC)}

\software{
SExtractor \citep{Bertin1996},
IRAF
}

\clearpage

\appendix

\section{Comparison of Lyman-alpha and UV SB profiles for each subsample}\label{sec:lyasbsep}

\begin{figure*}
\plottwo{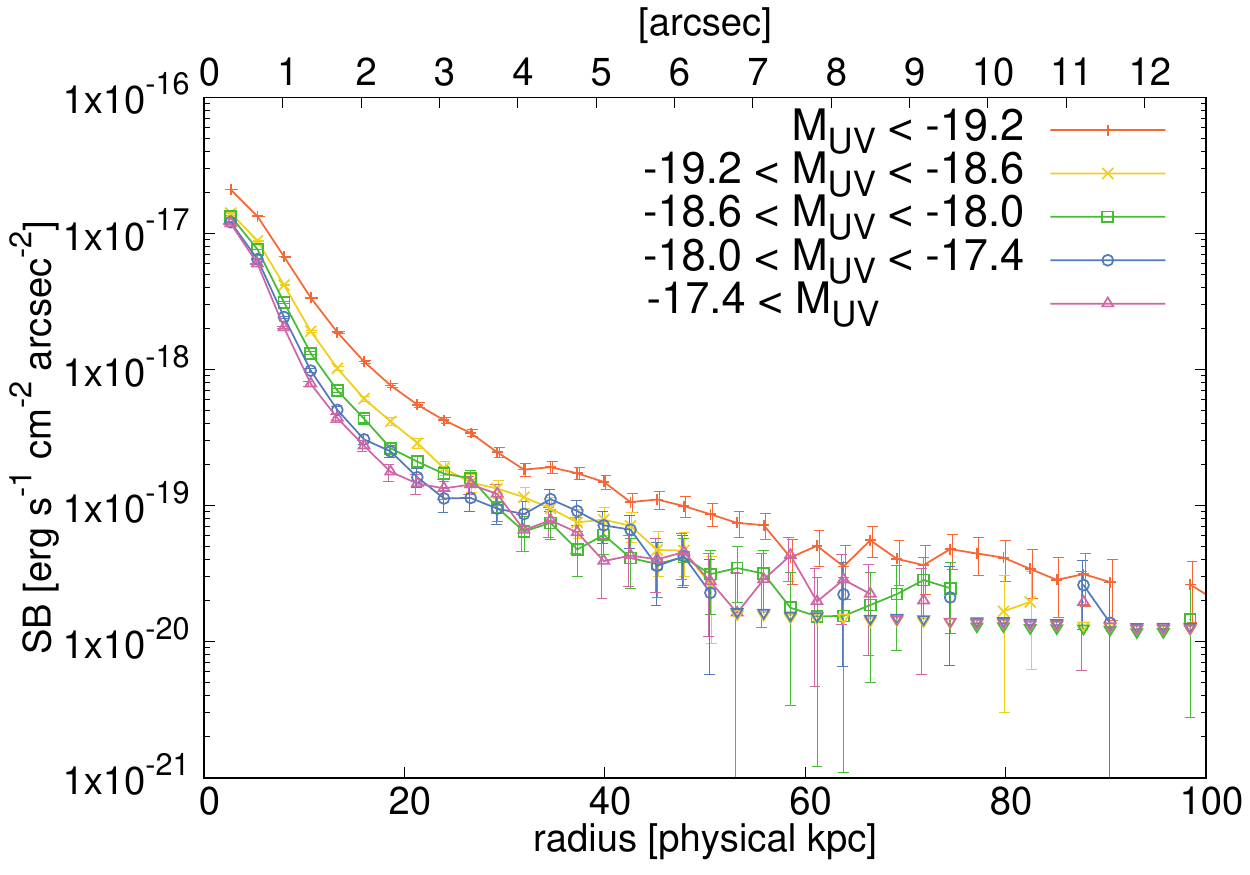}{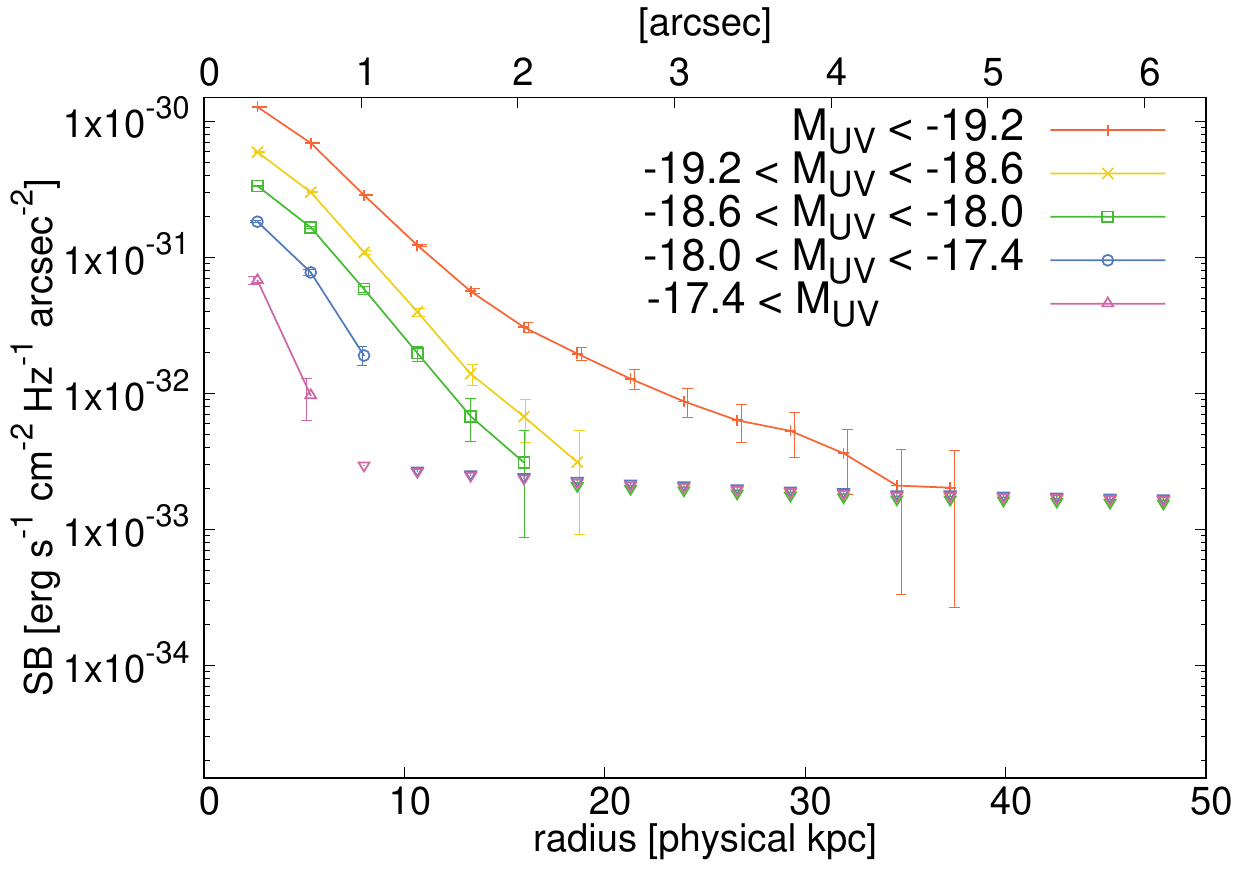}
\plottwo{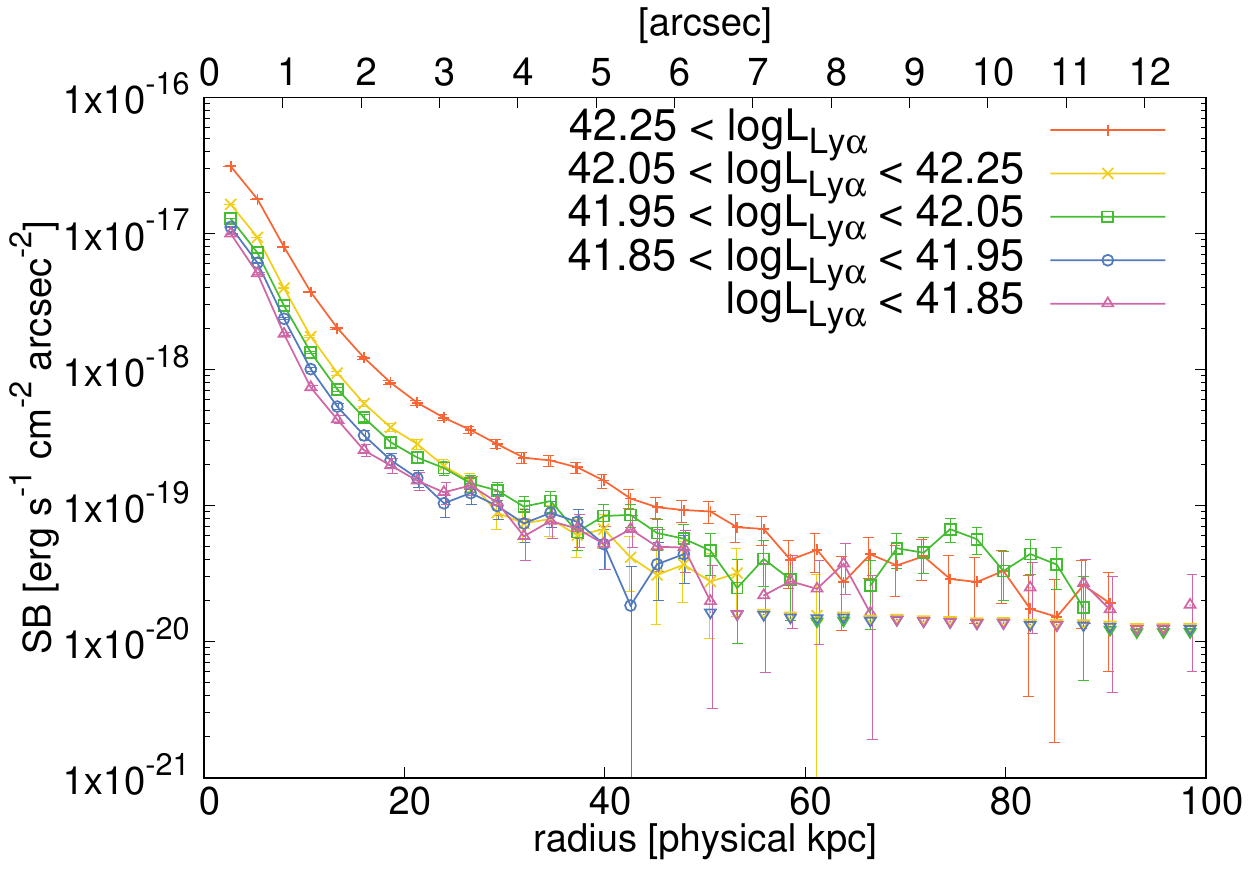}{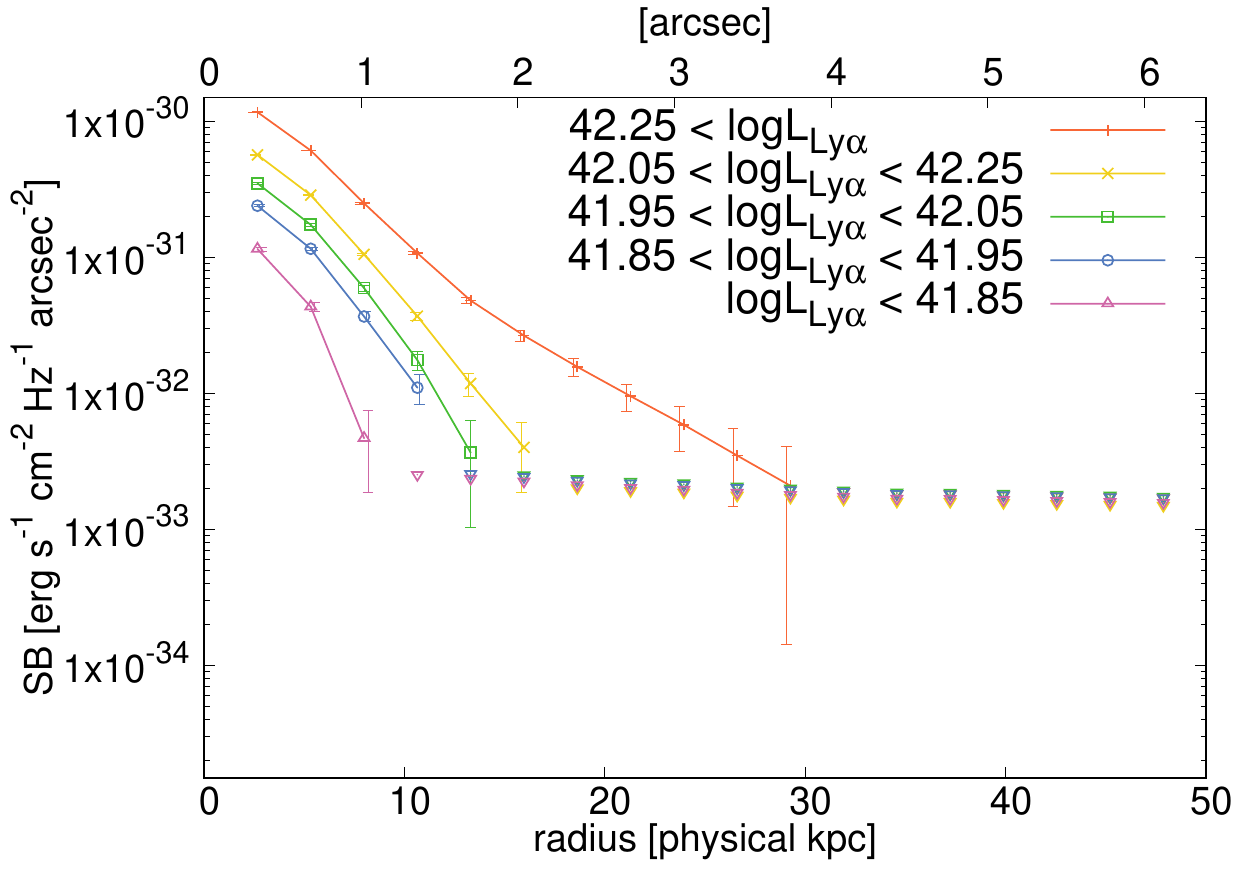}
\plottwo{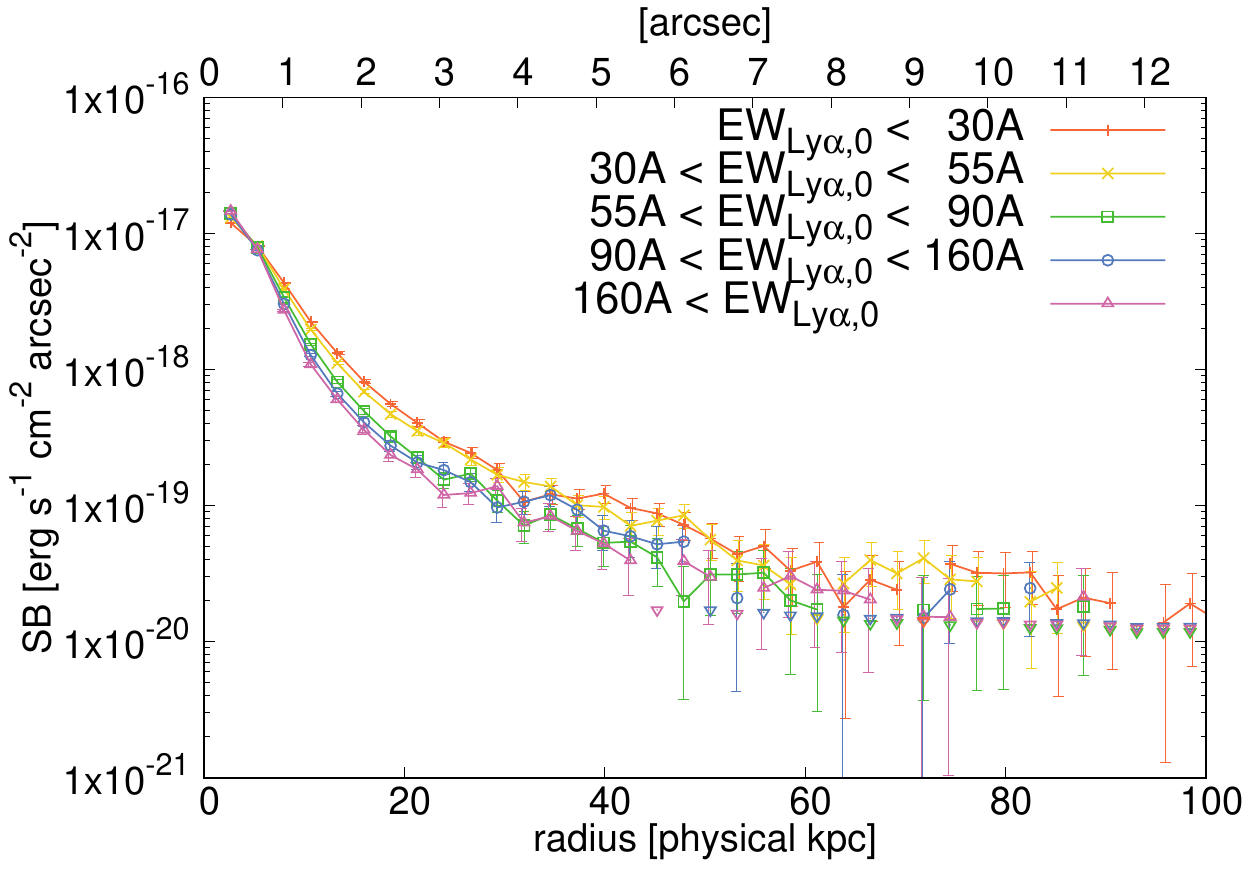}{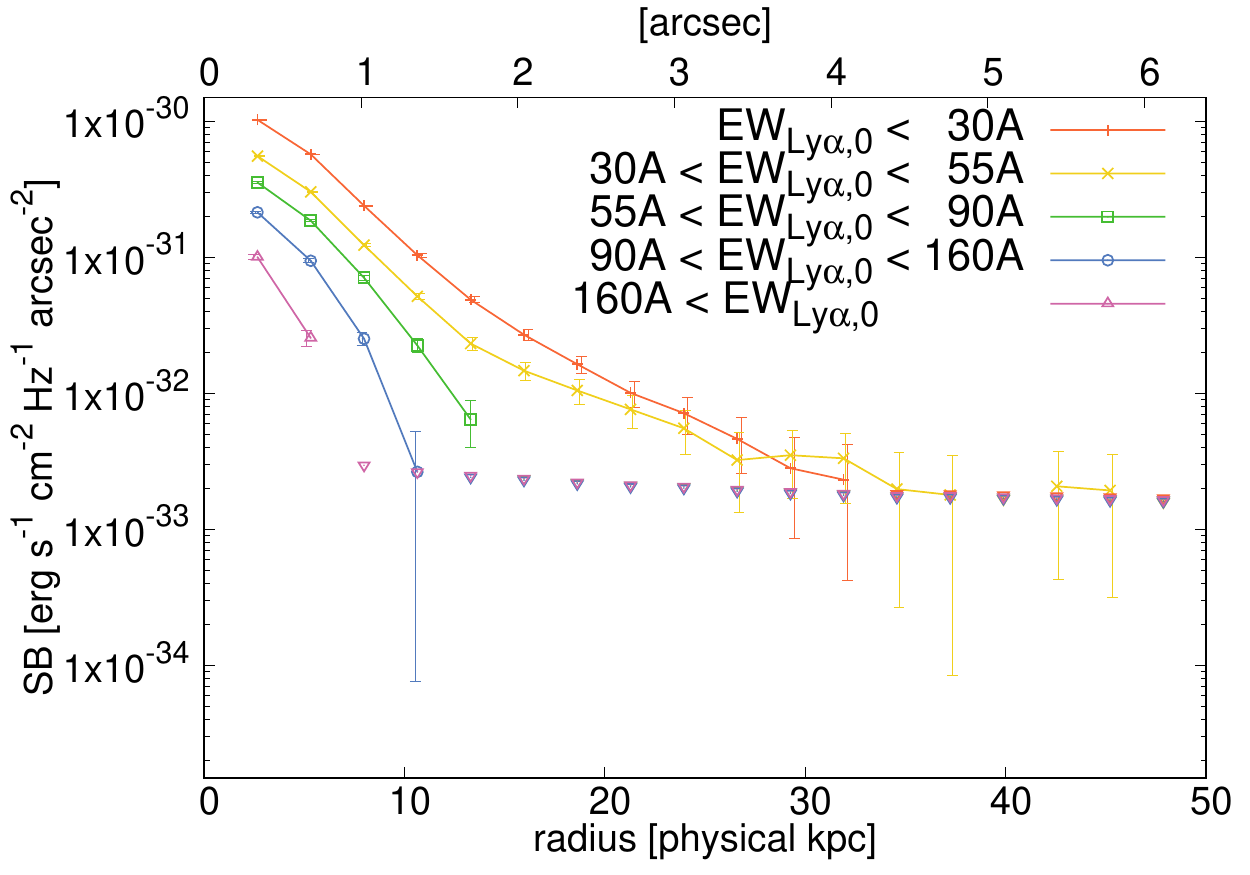}
\caption{Ly$\alpha$ SB profiles (Left) and UV continuum SB profiles (Right) derived by stacking analysis. From top to bottom, the LAE sample is grouped by their UV magnitude, Ly$\alpha$ luminosity, and Ly$\alpha$ equivalent width in the manner specified in Table \ref{tab:halostack}. The error bars are slightly shifted horizontally for display purpose. Downward triangles show 1$\sigma$ error levels after residual sky subtraction. }
\label{fig:Lyaprofs1}
\end{figure*}

\begin{figure*}
\plottwo{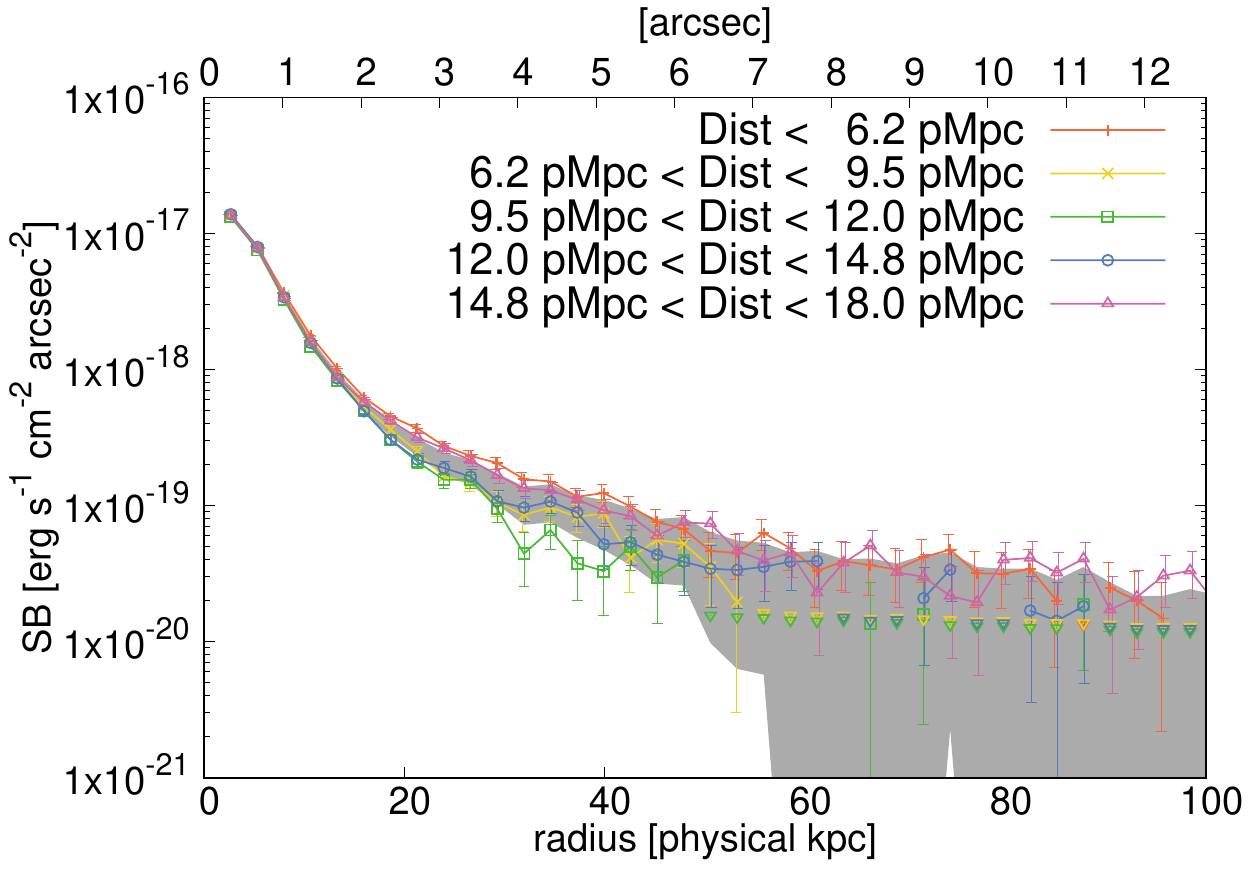}{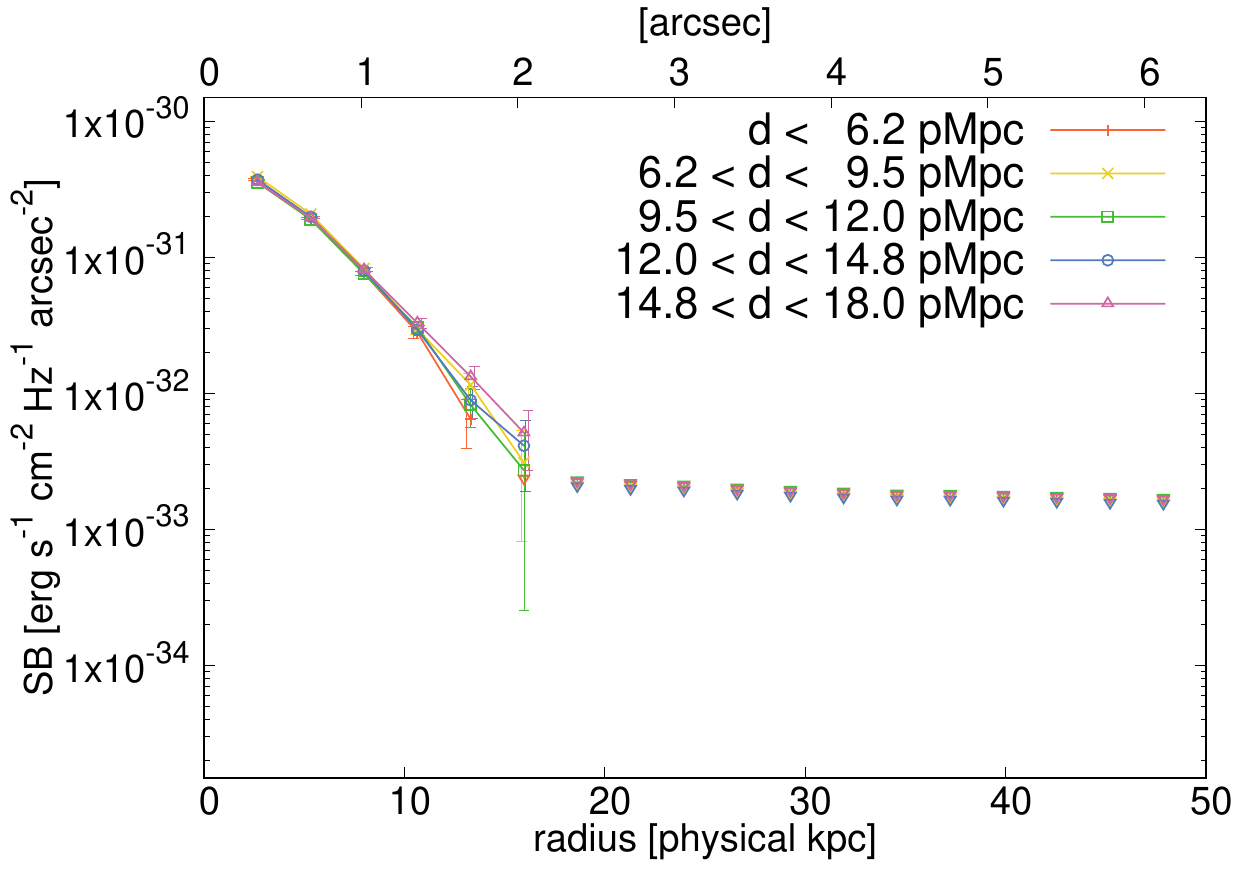}
\plottwo{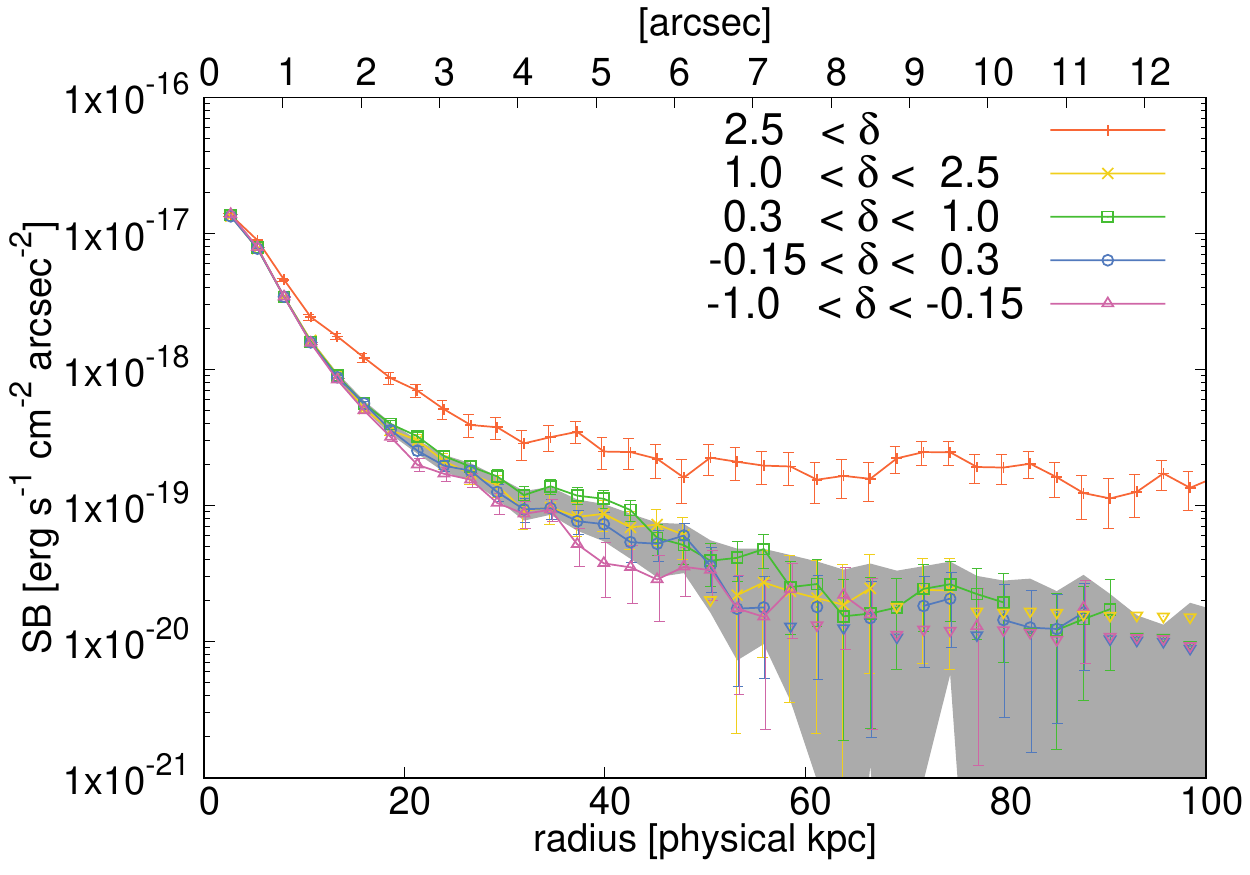}{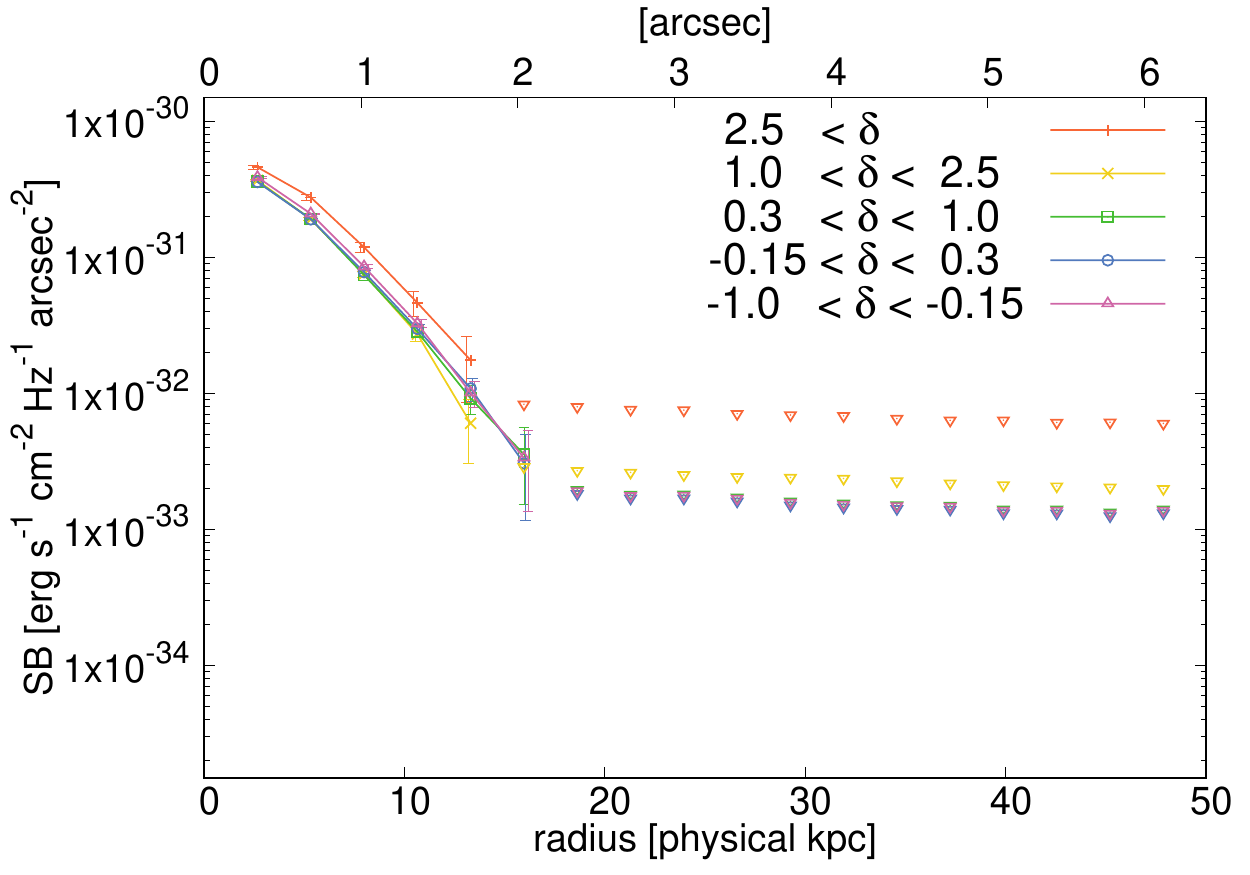}
\caption{Same as Figure \ref{fig:Lyaprofs1}, but here the LAEs are grouped by their projected distances from the HLQSO (Top) and local environment (Bottom) in the manner specified in Table \ref{tab:halostack}. The gray shaded regions in the Top (Bottom) panel shows 5th and 95th percentile of the Ly$\alpha$ SB distribution of stacked images created with randomly selected 700 (1000) LAEs. }
\label{fig:Lyaprofs2}
\end{figure*}

In Figures \ref{fig:Lyaprofs1} and \ref{fig:Lyaprofs2}, we show Ly$\alpha$ (left) and UV (right) SB profiles of each subsample in each row for easier comparison. 

In the top-left panel of Figure \ref{fig:Lyaprofs2}, the curves appear to deviate from each other beyond 2 arcsec, although the sensitivity at these angular separations is not very high. To check whether this difference is significant, we stacked Ly$\alpha$ images of 700 randomly selected LAEs (roughly corresponding to the number of LAEs in each bin of projected distance subsample) from the whole ($N=3490$) LAE sample and derived its Ly$\alpha$ SB profile. We repeated this 500 times and derived the 5th and 95th percentile of the SB distribution in each annulus. These are shown as a gray shaded region in Figure \ref{fig:Lyaprofs2}; in the bottom panel, we also show the SB distribution with 1000 randomly selected LAEs to see the difference between $\delta<1.0$ subsamples. The curves are almost within the shaded regions, suggesting that the difference is apparently marginal. 

\end{document}